\documentclass[oldversion]{aa} 
\usepackage{graphicx}
\usepackage{longtable}
\usepackage{txfonts}
\usepackage{rotating}
\usepackage{lscape}
\newcommand{\gsim}{\;\lower.6ex\hbox{$\sim$}\kern-7.75pt\raise.65ex\hbox{$>$}\;}
\newcommand{\lsim}{\;\lower.6ex\hbox{$\sim$}\kern-7.75pt\raise.65ex\hbox{$<$}\;}

\begin{document}
\title{The Na-O anticorrelation in horizontal branch stars. III. 47~Tuc and M~5
\thanks{Based on observations collected at 
ESO telescopes under programme 087.D-0230}
\fnmsep\thanks{
   Tables 2, 3, 4, 5 and 6 are only available in electronic form at the CDS via anonymous
   ftp to {\tt cdsarc.u-strasbg.fr} (130.79.128.5) or via
   {\tt http://cdsweb.u-strasbg.fr/cgi-bin/qcat?J/A+A/???/???}}
}

\author{
R.G. Gratton\inst{1},
S. Lucatello\inst{1},
A. Sollima\inst{1},
E. Carretta\inst{2},
A. Bragaglia\inst{2},
Y. Momany\inst{1,3},
V. D'Orazi\inst{1,4,5},
S. Cassisi\inst{6},
A. Pietrinferni\inst{6}
\and
M. Salaris\inst{7}}

\authorrunning{R.G. Gratton}
\titlerunning{Na-O in HB stars of 47~Tuc and M~5}

\offprints{R.G. Gratton, raffaele.gratton@oapd.inaf.it}

\institute{
INAF-Osservatorio Astronomico di Padova, Vicolo dell'Osservatorio 5, I-35122
 Padova, Italy
\and
INAF-Osservatorio Astronomico di Bologna, Via Ranzani 1, I-40127, Bologna, Italy
\and
European Southern Observatory, Alonso de Cordova 3107, Vitacura, Santiago, Chile 
\and
Department of Physics \& Astronomy, Macquarie University, Balaclava Rd., North Ryde, Sydney, NSW 2109, Australia
\and
Monash Centre for Astrophysics, School of Mathematical Sciences, Building 28, Monash University, VIC 3800, Australia
\and
INAF-Osservatorio Astronomico di Teramo, Via Collurania, Teramo, Italy
\and
Astrophysics Research Institute, Liverpool John Moores University, Twelve Quays House, Birkenhead, UK}

\date{}
\abstract{To check the impact of the multiple population scenario for
globular clusters on their horizontal branch (HB), we present an analysis of the 
composition of 110 red HB (RHB) stars in 47~Tucanae and of 61 blue HB (BHB) and 30 RHB 
stars in M~5. In 47~Tuc we found tight relations between the colours of the stars and their 
abundances of $p-$capture elements. This strongly supports the idea that the He content 
- which is expected to be closely correlated with the abundances of $p-$capture elements 
- is the third parameter (after overall metallicity and age) that determines the colour of 
HB stars. However, the range in He abundance must be small ($\Delta Y<0.03$) in 47~Tuc 
to reproduce our observations; this agrees with previous analyses. There is possibly 
a correlation between the abundances of $p-$\ and $n-$capture elements in 47~Tuc. If 
confirmed, this might suggest that asymptotic giant branch stars of moderate mass contributed to the gas from 
which second-generation stars formed. Considering the selection effects in our 
sample (which does not include stars warmer than 11000~K and RR Lyrae variables, which were 
excluded because we could not obtain accurate abundances with the adopted observing 
procedure) is important to understand our results for M~5. In this case, we find that, as 
expected, RHB stars are Na-poor and O-rich, and likely belong to the primordial 
population. There is a clear correlation of the [Na/O] ratio and N abundance with colour 
along the BHB. A derivation of the He abundance for these stars yields a low value of 
$Y=0.22\pm 0.03$. This is expected because HB stars of a putative He-rich population in 
this cluster should be warmer than 11000~K, and would accordingly not have been sampled by 
our analysis. However, we need some additional source of scatter in the total mass loss 
of stars climbing up the red giant branch to reproduce our results for M~5. Finally, 
we found a C-star on the HB of 47~Tuc and a Ba-rich, fast-rotating, likely 
binary star on the HB of M~5. These stars are among the brightest and coolest HB stars.}

\maketitle

\section{Introduction}

Core He-burning stars in globular clusters (GC) are distributed along the so-called
horizontal branch (HB) in the colour-magnitude diagram (CMD). It is well known that
this distribution may be very different from cluster to cluster. The precise 
location of an individual star along the HB changes due to evolution; however, 
most of the lifetime on the HB is spent close to the zero age horizontal branch 
(ZAHB) location. Hence, the colour distribution of stars along the HB largely, though not
uniquely, reflects their ZAHB distribution. In turn, several parameters may
affect this distribution: total mass (itself related to various parameters,
such as age, metallicity, and He content), core mass, overall metallicity, CNO/Fe
ratio, etc. (see e.g. Catelan \& De Freitas Pacheco 1995).
Some of these quantities might be related to very poorly known
parameters, such as core rotation or details of the mass loss. In principle, this allows 
to use HB stars as a very sensitive diagnostic for quantities otherwise difficult 
to determine. Separating these effects is difficult: progress
has been quite slow and many facts still need an adequate explanation.

It has been clear for nearly half a century that the main parameter is related to the
overall metal content (Sandage \& Wallerstein 1960; Faulkner 1966). For a long time, 
age has been considered the most likely candidate for the second parameter (see Lee et al. 1994). 
The small age difference between GCs implied that confirming this assumption is difficult and
it was then achieved only recently (Dotter et al. 2010; Gratton et al. 2010). 
Even more recently, it has become 
clear that star-to-star He abundance variations within a cluster, related to the presence 
of multiple populations in most if not all GCs, are the most likely third parameter. This 
is because He-rich stars (known to be present in GCs: e.g. Piotto et al. 2005) evolve faster on 
the main sequence and then their progeny on the HB is less massive than that of He-poor stars 
(Norris et al. 1981; D'Antona et al. 2002; D'Antona \& Caloi 2004). There are many 
circumstantial facts supporting this assertion (Carretta et al. 2009a; Gratton et al. 2010). 
Direct confirmation may come from spectroscopic analysis of the chemical composition of stars 
along the HB. The surface He abundance of evolved stars is indeed slightly higher than the 
original value due to the first dredge up (see e.g. Sweigart 1987); however, this difference is 
small ($\sim 0.015$ in Y). Unfortunately, determining the initial He content of HB stars 
is possible only for a limited range of temperatures, 
because the surface composition of stars with $T_{\rm eff}>11,000$~K is strongly modified 
by sedimentation effects (see e.g. Behr et al. 1999) and He lines are vanishingly weak in 
cool stars. Furthermore, determinations need to be accurate to be really meaningful. While 
some determinations of He abundances of adequate precision and accuracy have been made for 
stars with $9000<T_{\rm eff}<11,000$~K (Villanova et al. 2009, 2012), this temperature 
limitation is severe. It is much easier to determine the abundances of other elements 
(including e.g. Na and O) that are thought to be correlated with He in the multiple 
population scenario (see reviews by Gratton et al. 2004, 2012a). They can be determined 
for all stars cooler than $T_{\rm eff}<11,000$~K; this makes the test useful for a quite 
large number of GCs. Marino et al. (2011) presented such a determination for M~4, showing 
that Na-poor/O-rich stars lie on the RHB and Na-rich/O-poor are on the BHB of this GC, as 
expected if the colour distribution of stars along the HB is mainly determined by He-abundance
variations. To extend this result to a wider sample of GCs, we began a systematic survey of 
the abundances of O and Na (and other relevant elements) in HB stars of a few GCs. Clusters 
were selected to have a large number of stars cooler than $11,000$~K, but with 
a wide range of properties. We previously presented data for NGC~2808 (Gratton et al. 2011, 
Paper I) and NGC~1851 (Gratton et al. 2012b, Paper II). In this paper we present 
the analysis of 47~Tuc (=NGC~104) and M~5 (=NGC~5904). Results for a few other 
GCs (NGC~6388, NGC~6656=M~22, NGC~6723) will be presented in future papers.

Both 47~Tuc and M~5 are well-studied GCs. The Na-O anticorrelation for a
large number of red giant branch (RGB) stars of these clusters is presented in Carretta et al.
(2009a, 2009b). Both clusters have an Na-O anticorrelation, which for 47~Tuc is short  
for such a massive GC, while M~5 is a more typical massive
GC with some very O-poor stars. 47~Tuc is the classical example of metal-rich GCs 
with red HB. A few authors (see Mohler et al. 2000 and references therein) found 
a few blue HB stars in 47~Tuc. Their origin is not known, but they likely are not 
connected to an extremely rich He population but rather have a different
origin, e.g. they may be stars that have lost an abnormal amount of mass in mass
transfer episodes in binary systems. We have not observed any of these stars in our
programme, and we will not discuss them any further. Among the many photometric analyses 
of the HB, the most interesting to 
us are the very recent works by Di Criscienzo et al. (2010) and Nataf et al. (2011), 
who found evidence for a spread in He by 0.02~per cent in mass, and by Milone et al.
(2012), who confirmed this result and showed that the different populations
present in this cluster can be traced from the main sequence up to the HB.
Low-resolution spectra of HB stars have been analysed by Norris \& Freeman (1982),
who found evidence for a C-N anticorrelation. While the number of stars observed
was quite small and only qualitative information on composition were obtained, it 
has been recently recognized that CN-poor HB stars in this sample are redder and 
fainter than CN-rich ones (Milone et al. 2012), as expected if He determines the
colour of HB stars. A very similar result was obtained by Briley (1997): see in
particular his impressive Figure~8. In addition, Nataf et al. (2011) noticed that there is a
gradient in the colour of HB stars within the cluster, bluer stars being more
centrally concentrated than redder ones. These authors interpreted this finding in the
framework of a multiple populations scenario as evidence that second-generation
stars are more centrally concentrated than first-generation ones. Again, a similar
result was previously obtained by Briley (1997). These studies 
essentially confirm that the HB of 47~Tuc contains different populations; however, a 
high-dispersion spectroscopic study of a much larger sample of stars is required to 
confirm the suggested correlation between the location of stars on the 
HB and their composition. 

The extended HB of M~5 has been studied photometrically. After the pioneering study
by Greenstein \& Munch (1966), who noticed the weakness of He lines in warm blue HB
stars in this and other globular clusters, Peterson (1983) determined radial and rotational 
velocities for seven blue HB (BHB) stars; Crocker et al. (1988) determined surface gravities 
for 12 BHB stars in agreement with the values expected from HB models; and Lai 
et al. (2011) examined two red HB (RHB) stars finding that they are O-rich 
and moderately Na-poor. The observational material on which these papers are based
is not sufficient to discuss the relation between HB and the Na-O anticorrelation.

The outline of this paper is as follows: in Section 2 we present observational
data. Details on the analysis - which is very similar to those already presented
in Papers I and II - are given in Section 3. Results are presented and discussed 
in Section 4, while conclusions are drawn in Section 5.

\begin{center}
\begin{figure}
\includegraphics[width=8.8cm]{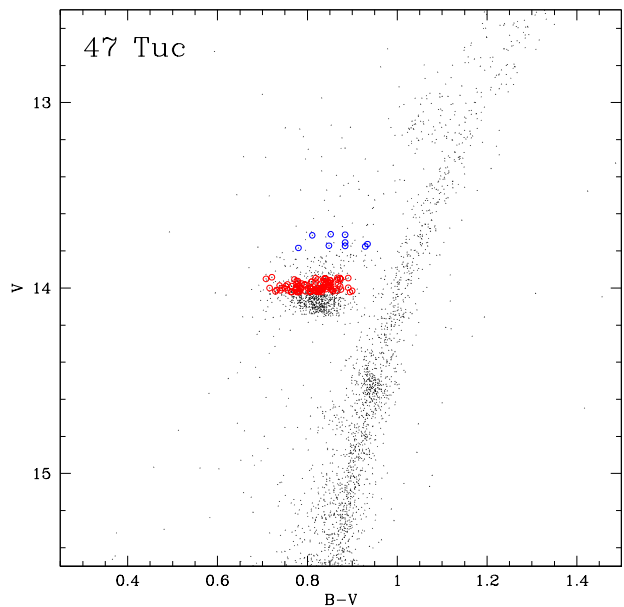}
\caption{Colour-magnitude diagram of 47~Tuc. Circles are the stars
analysed in this paper: red symbols are our faint stars group, while the blue ones are
the bright stars group. Dots are stars not observed in this paper. }
\label{f:fig1}
\end{figure}
\end{center}

\begin{center}
\begin{figure}
\includegraphics[width=8.8cm]{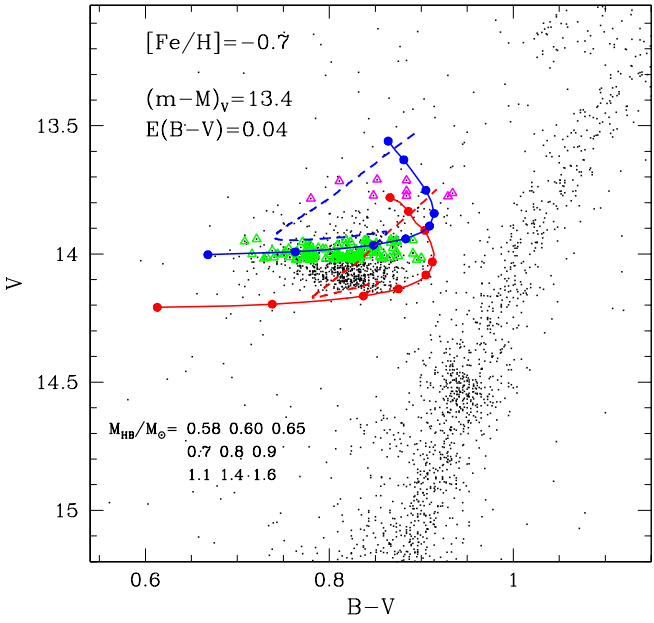}
\caption{Colour-magnitude diagram of 47~Tuc. Triangles are the stars
analysed in this paper: green symbols are our faint stars group, while the magenta ones are
the bright stars group. Dots are stars not observed in this paper. For comparison, location
of ZAHB stars of different masses and He abundances (Y=0.25: red dots and solid line; Y=0.28:
blue dots and solid line), the evolutionary tracks off the ZAHB for stars
of $0.65~M_\odot$\ (dashed lines) are also plotted.}
\label{f:bright}
\end{figure}
\end{center}

\begin{center}
\begin{figure}
\includegraphics[width=8.8cm]{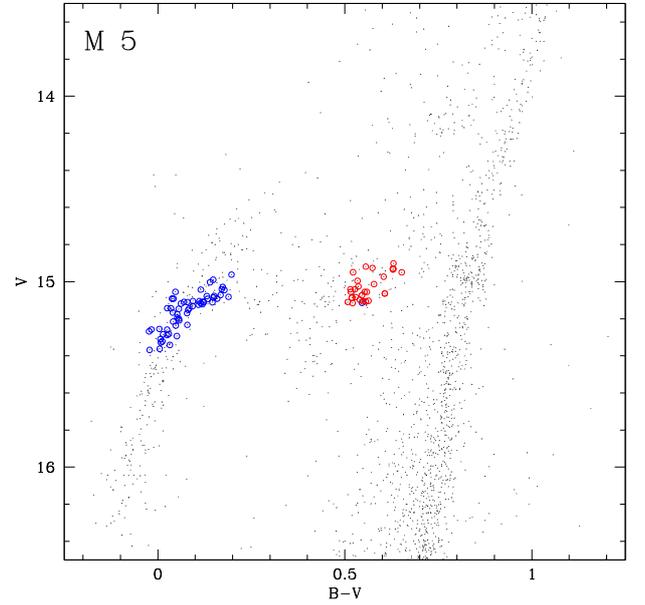}
\caption{Colour-magnitude diagram of M~5. Red circles are RHB stars, blue
circles are BHB stars. Dots are stars not observed
in this paper.}
\label{f:fig2}
\end{figure}
\end{center}

\begin{center}
\begin{figure}
\includegraphics[width=8.8cm]{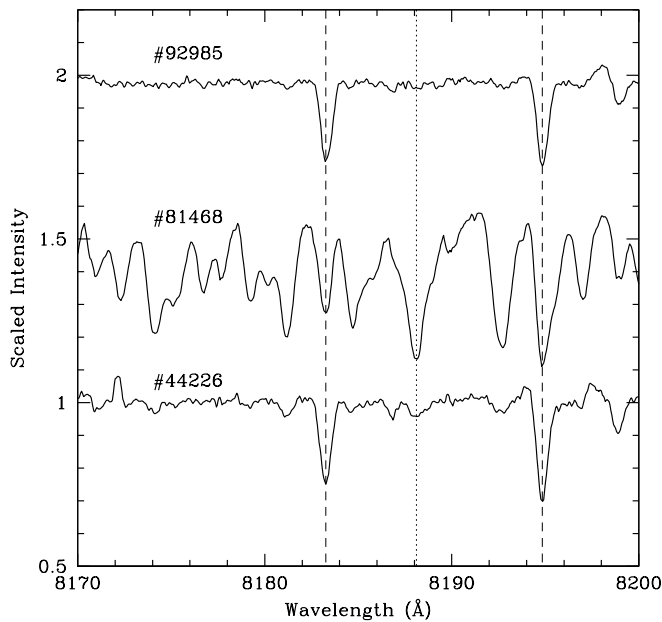}
\caption{Portion of the spectra of three RHB stars of 47~Tuc (\#92985: upper row; 
\#81468: middle row; \#44226: lower row). The first two spectra are offset
vertically for clarity. These spectra have been shifted at zero radial velocity; 
telluric features are excised following the technique described in the text. 
Dashed lines mark the Na{\ts I} doublet at 8183-94~\AA; the dotted line marks the 
strongest CN line in this spectral region (many other CN lines are present). Note
that \#81468 is a C-star, discussed in Section 4.8. }
\label{f:fig3}
\end{figure}
\end{center}

\begin{table}[htb]
\centering
\caption[]{Observing log}
\begin{tabular}{ccccc}
\hline
NGC  & Grating &    Date     &  UT      &  Exp. Time \\
     & HR      &             &          & (s) \\
\hline            
5904 & 12 &  2011-05-13 & 01:44:33 & 1900 \\
5904 & 19 &  2011-05-20 & 02:50:18 & 3675 \\
104  & 12 &  2011-06-25 & 08:45:37 & 1400 \\
104  & 19 &  2011-06-25 & 09:22:19 & 2300 \\
\hline
\end{tabular}
\label{t:tab0}
\end{table}

\begin{table*}[htb]
\centering
\caption[]{Basic data for programme stars in 47~Tuc (complete table available only electronically at CDS)}
\begin{scriptsize}
\begin{tabular}{cccccccccccccc}
\hline
Star & RA & dec & S/N & S/N & $V_r$ & FWHM & $B$ & $V$ &  $K$   &$(B-V)_o$&$(V-K)_o$&$T_{\rm eff}$&$\log{g}$\\
     &(J2000)&(J2000)& HR12 & HR19 & (km$\ts $s$^{-1}$)&(km$\ts $s$^{-1}$)& (mag) & (mag) & (mag) & (mag) & (mag) & (K) &  \\ 
\hline
\multicolumn{14}{c}{Bright stars}\\
\hline
~16399	& 0~23~38.676 & -71~56~00.82 & 49 & 90 &-25.9 & 33.0 &14.639 & 13.755 & 11.474 & 0.844 & 2.172 & 4969 & 2.15 \\
~41553	& 0~23~28.770 & -72~08~53.34 & 70 & 57 &-22.0 & 33.6 &14.597 & 13.713 & 11.494 & 0.844 & 2.110 & 5003 & 2.15 \\
~74577	& 0~24~47.123 & -72~08~11.09 & 68 &103 &-18.0 & 34.5 &14.620 & 13.772 & 11.693 & 0.808 & 1.970 & 5124 & 2.22 \\
~81468	& 0~23~58.824 & -72~05~59.95 & 68 &    &-19.6 & 33.3 &14.697 & 13.763 & 11.492 & 0.894 & 2.162 & 4919 & 2.13 \\
~81868	& 0~23~47.838 & -72~05~53.32 & 60 &    &-29.4 & 37.2 &14.564 & 13.784 &	       & 0.740 &       & 5263 & 2.29 \\
\hline
\end{tabular}
\end{scriptsize}
\label{t:tab1}
\end{table*}

\begin{table*}[htb]
\centering
\caption[]{Basic data for programme stars in M~5 (complete table available only electronically at CDS)}
\begin{scriptsize}
\begin{tabular}{ccccccccccccccc}
\hline
Star & RA & Dec & S/N & S/N & $V_r$ & FWHM & $B$ & $V$ &  $K$   &$(B-V)_o$&$(V-K)_o$&$T_{\rm eff}$&$\log{g}$&$v_t$ \\
     &(J2000)&(J2000)& HR12 & HR19 &(km$\ts $s$^{-1}$)&(km$\ts $s$^{-1}$)& (mag) & (mag) & (mag) & (mag) & (mag) & (K) & & (km$\ts $s$^{-1}$) \\ 
\hline
\multicolumn{15}{c}{Blue HB stars}\\
\hline
~5037	& 15~18~07.726 & 1~56~56.90 & 39 & 37 & 52.8& 33.6 &15.179 & 15.144 & 15.068 & 0.005 & -0.006 & 9082 & 3.28 & 2.0 \\
~6864	& 15~18~11.488 & 2~03~06.40 & 36 & 42 & 47.6& 35.4 &15.373 & 15.355 & 15.153 &-0.012 & 	0.120 & 9392 & 3.40 & 2.0 \\
~7304	& 15~17~59.026 & 2~04~02.15 & 40 & 64 & 49.6& 30.0 &15.057 & 14.936 & 14.399 & 0.091 & 	0.455 & 8167 & 3.06 & 2.4 \\
~7432	& 15~17~58.620 & 2~04~17.53 & 35 & 52 & 53.9& 29.8 &15.300 & 15.265 & 14.773 & 0.005 & 	0.410 & 9082 & 3.32 & 2.0 \\
~8562	& 15~17~44.520 & 2~06~57.04 & 33 & 37 & 52.6& 27.8 &15.176 & 15.169 & 14.802 &-0.023 & 	0.285 & 9622 & 3.35 & 2.0 \\
\hline
\end{tabular}
\end{scriptsize}
\label{t:tab2}
\end{table*}

\section{Observations}

We acquired spectra of 110 RHB stars in 47~Tuc (nine of them much brighter than the remaining
ones; they are either objects in very late phases of HB evolution or massive objects resulting
from evolution in mass transfer binaries), and of 61 BHB and 30 RHB stars 
(including 12 bright and likely evolved stars) in M~5 using the GIRAFFE fibre-fed 
spectrograph at the VLT (Pasquini et al. 2004). Three additional RHB candidates 
in M5 turned out to be field interlopers with a clearly deviating radial velocity and 
much higher metallicity than cluster members. The spectral resolution is $R=18700$ for grating
R12 (wavelength range 5821-6146~\AA) and $R=13900$\ for grating HR19A (wavelength range 
7745-8335~\AA). Our program was executed in service mode. 
Single observations were obtained with the individual gratings. The log of observations 
is given in Table~\ref{t:tab0}. Figures~\ref{f:fig1} and ~\ref{f:fig2} show the location 
of the programme stars on the colour magnitude diagrams of 47~Tuc and M~5, respectively. 
Our ground-based photometric catalogue (see Momany et al. 2004) consists of $UBVI$\ 
observations obtained with the Wide-Field Imager (WFI) mounted on the 2.2m  ESO-MPI 
telescope (La Silla, Chile). The WFI catalogue has a total field of view of 
$34^{\prime}\times~33^{\prime}$.  Photometric data for the programme stars are listed 
in Table~\ref{t:tab1}. Note that at variance with the other colours, the $U$\ photometry 
is not calibrated. The $K$\ magnitudes are from the 2MASS point source catalogue 
(Skrutskie et al. 2006). 

Stars in 47~Tuc are spread over the whole range in colour of the RHB ($0.708<(B-V)<0.891$) 
but in two restricted ranges in magnitude: nine bright objects ($13.7<V<13.8$) and a much 
larger group of fainter ones ($13.94<V<14.03$). This selection in magnitude is not 
representative of the distribution in magnitude of the whole sample of RHB stars. The 
brighter group of objects includes either objects in the later phases of HB evolution or
more massive objects resulting from the evolution of binary systems (see Section 4.8).
Figure~\ref{f:bright} compares the location of the programme stars with ZAHB
models for different masses and He abundances, as well with evolutionary tracks off the
ZAHB for stars of 0.65~$M_\odot$. The group of bright stars is represented
by magenta triangles. As can be seen, they may either be ZAHB stars with masses
in the range 1.1-1.6~$M_\odot$, or stars of $\sim 0.65~M_\odot$ in later
evolutionary phase. In the first case, they may result from the evolution of blue
straggler stars. As to the fainter group, since the ZAHB of 47~Tuc is slightly tilted towards the red (Di 
Criscienzo et al. 2010; Nataf et al. 2011), our sample includes blue objects close
to the ZAHB as well as slightly evolved objects in the redder region. Since evolution 
off the ZAHB becomes faster with increasing luminosity, reconstruction of the true
distribution of stars with different chemical composition requires a comparison with
synthetic HBs. However, He-rich (and the Na-rich and O-poor) stars are likely over-represented 
in our sample. All BHB stars we observed in M~5 are cooler than the Grundahl et al. 
(1999) $u-$jump, and could be used in our analysis.

All programme stars were chosen to be free from any companion closer 
than 2 arcsec and brighter than $V+2$~mag, where $V$ is the target magnitude. 
The remaining fibres were used to acquire sky spectra. The median spectra from these 
last fibres were subtracted from those used for the stars. 

We used two spectral configurations: HR12 and HR19, providing high-resolution spectra including 
the strongest features of O\ts{I} (the IR triplet at 7771-74~\AA) and Na\ts{I} (the 
resonance D doublet at 5890-96~\AA, as well as the subordinate strong doublet at 
8183-94~\AA) accessible from ground and the only ones that might be used to determine 
O and Na abundances without a prohibitively long observing time. A few lines of N, Mg, 
Al, Si, Ca, Fe, and Ba, as well as a number of CN lines, are also included in the 
selected observing ranges.

The signal-to-noise (S/N) ratio of the spectra is typically 
$\sim 50-60$ and $\sim 80$ for 47~Tuc stars, and $\sim 40-50$ and $\sim 50-70$ , for 
those in M~5 (for each cluster, the first number is for HR12 and the second one is for 
HR19). The spectra were reduced by the ESO personnel using the ESO FLAMES GIRAFFE 
pipeline version 2.8.7. Sky subtraction, translation to rest-frame and continuum tracing 
were performed within IRAF \footnote{IRAF is distributed by the National Optical Astronomical 
Observatory, which are operated by the Association of Universities for Research in 
Astronomy, under contract with the National Science Foundation}. Telluric absorption lines 
were removed from the longest wavelength spectra by dividing them for the average 
spectrum of the warmer HB stars (those with $T_{\rm eff}>11,500$~K) in NGC~1851,
which are essentially featureless in the spectral region of interest.
A few very high excitation lines are present in this spectral
region, e.g. the OI and NI lines. However, radial velocity of stars in NGC~1851 
is so much different from those of the stars in M5 and 47 Tuc that there is no danger of 
over-positioning the lines. This step
required some adaptation to achieve a similar telluric line strength because the spectra 
of NGC~1851 HB stars were not acquired with the same airmass as the programme ones. The 
excision of the telluric lines was efficient, however. Examples of spectra are 
shown in Figure~\ref{f:fig3}.

\subsection{Membership and radial velocities}

A comparison with the Galactic model by Robin et al. 
(2003) indicates that RHB stars of 47~Tuc have a membership probability $P>96$\%
(90\% of the stars with $P>99$\%) and there is a 41\% probability that all stars 
are cluster members, once colours, magnitudes, stellar density, and radial velocities 
are taken into account. 

While such data were not available for the programme stars in
47~Tuc, Cudworth (1979) provided membership probability for bright stars in M5 based on proper
motions. There are 42 stars in common with our sample. For 39 stars, the membership
probabilities is $>95$\%. The three remaining cases (two BHB stars: \#31865 and \#32273;
and an RHB one: \#40293) have a membership probability from proper motions between 1 and 18\%. However, they 
are also likely cluster members. Indeed, the models by Robin et al. (2003) 
indicate that we do not expect any contamination for the BHB of M~5. Although
some contamination is possible for RHB stars, the probability that all RHB stars are 
cluster members is 80\% , and \#40293 is not distinguishable from M~5 stars for any other 
property. On the other hand, three stars with colours and magnitudes similar to those
on the RHB of M5 turned out to have discrepant radial velocities and metal abundances,
and were not considered to be members. They all lie far from the cluster centre at distances
larger than 8~arcmin, to be compared with half-light and tidal radii of 1.77 and 24
arcmin, respectively (Harris, 1996; we used the latest available version at
URL www.physics.mcmaster.ca/Globular.html). They were omitted from our analysis and tables.

Radial velocities $V_r$\ were obtained using the $fxcor$\ IRAF routine with suitable templates.
They are listed in Tables~\ref{t:tab1} and ~\ref{t:tab2} for 47~Tuc and M~5, respectively.
We obtained $V_r=-19.6\pm 0.9$~km$\ts $s$^{-1}$ (r.m.s of 9.5~km$\ts $s$^{-1}$) from 110 
stars in 47~Tuc. For comparison, the value listed by Harris (1996) is $-18.0\pm 0.1$~km$\ts $s$^{-1}$,
with a central velocity dispersion of 11.0~km$\ts $s$^{-1}$. For M~5, we obtained averages of 
$53.0\pm 0.4$~km$\ts $s$^{-1}$ (r.m.s of 3.5~km$\ts $s$^{-1}$) from 61 BHB stars and of $51.6\pm 0.9$~km$\ts $s$^{-1}$
(r.m.s of 5.0~km$\ts $s$^{-1}$) from 30 RHB stars. Harris' value is $53.2\pm 0.4$~km$\ts $s$^{-1}$, with a
central velocity dispersion of 5.5~km$\ts $s$^{-1}$. The overall agreement is excellent, although it
is possible that radial velocities for RHB stars have a small ($\sim 1$~km$\ts $s$^{-1}$) negative offset
that might be due to the template used.

%
%

\begin{table*}[htb]
\centering
\caption[]{Abundances of Fe, O, Na, and Al in HB stars of 47~Tuc (complete table available only electronically at CDS)}
\begin{scriptsize}
\begin{tabular}{cccccccccccccccccc}
\hline
Star &\multicolumn{3}{c}{$[$Fe/H$]_I$}&
      \multicolumn{3}{c}{$[$Fe/H$]_II$}&
      [N/Fe]&
      \multicolumn{3}{c}{$[$O/Fe$]$}&
      \multicolumn{3}{c}{$[$Na/Fe$]$}&
      [Mg/Fe]&
      \multicolumn{3}{c}{$[$Al/Fe$]$}\\
& n & $<>$ & rms &    
  n & $<>$ & rms &    
& n & $<>$ & rms &    
  n & $<>$ & rms &    
  n & $<>$ & rms \\   
\hline
\multicolumn{18}{c}{Bright stars}\\
\hline
~16399 & 33 & -0.62 & 0.20 & 3 & -0.93 & 0.21 & 1.72 & 3 & 	0.20 & 0.08	& 4 & 0.24 & 0.05 & 0.39 & 2 &  0.02 & 0.08 \\
~41553 & 31 & -0.71 & 0.29 & 3 & -0.78 & 0.17 & 1.68 & 3 & 	0.09 & 0.11 & 4 & 0.55 & 0.02 & 0.36 & 2 &  0.19 & 0.24 \\
~74577 & 31 & -0.69 & 0.18 & 3 & -0.73 & 0.10 & 1.53 & 3 & -0.13 & 0.12 & 4 & 0.63 & 0.10 & 0.41 & 2 &  0.29 & 0.06 \\
~81468 & 21 & -1.04 & 0.28 & 3 & -0.59 & 0.24 &      & 3 & 	0.50 & 0.16 & 4 & 0.58 & 0.17 & 	 & 1 & -0.02 &      \\
~81868 & 21 & -0.87 & 0.17 & 2 & -1.00 & 0.02 & 	 & 	 & 	     &      & 2 & 0.60 & 0.02 & 	 & 	 &       &      \\
\hline
\end{tabular}
\end{scriptsize}
\label{t:tab3}
\end{table*}

\begin{table*}[htb]
\centering
\caption[]{Abundances of Si, Ca, Ti, V, Mn, Ni, and Ba in HB stars of 47~Tuc (complete table available only electronically at CDS)}
\begin{scriptsize}
\begin{tabular}{cccccccccccccccccccc}
\hline
Star &\multicolumn{3}{c}{$[$Si/Fe$]$}&
      \multicolumn{3}{c}{$[$Ca/Fe$]$}&
      \multicolumn{3}{c}{$[$Ti/Fe$]$}&
      \multicolumn{3}{c}{$[$V/Fe$]$}&
      \multicolumn{3}{c}{$[$Mn/Fe$]$}&
      \multicolumn{3}{c}{$[$Ni/Fe$]$}&
$[$Ba/Fe$]$\\
& n & $<>$ & rms &    
  n & $<>$ & rms &    
  n & $<>$ & rms &    
  n & $<>$ & rms &    
  n & $<>$ & rms &    
  n & $<>$ & rms &    
\\
\hline
\multicolumn{20}{c}{Bright stars}\\
\hline
~16399 & 4 & 0.26 & 0.19 & 3 & 0.38 & 0.17 & 5 & 0.15 & 0.17 & 3 &-0.10 & 0.01 & 3 & -0.37 & 0.23 & 5 &-0.05 & 0.28 & 0.18 \\ 
~41553 & 4 & 0.42 & 0.16 & 3 & 0.49 & 0.26 & 5 & 0.22 & 0.25 & 3 &-0.01 & 0.06 & 3 & -0.31 & 0.17 & 5 & 0.09 & 0.22 & 0.28 \\ 
~74577 & 4 & 0.24 & 0.27 & 3 & 0.43 & 0.35 & 5 & 0.13 & 0.13 & 3 &-0.03 & 0.16 & 3 & -0.37 & 0.09 & 5 & 0.04 & 0.23 & 0.49 \\ 
~81468 & 4 &      &      & 3 & 0.47 & 0.27 & 5 & 0.39 & 0.25 & 3 &      & 0.32 & 3 & -0.41 & 0.32 & 5 & 0.14 & 0.21 & 1.50 \\ 
~81868 & 2 & 0.27 &	0.06 & 3 & 0.66 & 0.08 & 4 & 0.11 & 0.10 & 2 & 0.00 & 0.04 & 3 & -0.48 & 0.03 & 5 &-0.08 & 0.15 & 0.49 \\ 
\hline
\end{tabular}
\end{scriptsize}
\label{t:tab4}
\end{table*}

\begin{table*}[htb]
\centering
\caption[]{Elemental abundances in HB stars of M~5 (complete table available only electronically at CDS)}
\begin{scriptsize}
\begin{tabular}{cccccccccccccccc}
\hline
Star &\multicolumn{3}{c}{$[$Fe/H$]_I$}&
      \multicolumn{3}{c}{$[$Fe/H$]_II$}&
$[$N/Fe$]$&
$[$O/Fe$]$&
$[$Na/Fe$]$&
$[$Mg/Fe$]$&
$[$Si/Fe$]$&
$[$Ca/Fe$]$&
$[$Mn/Fe$]$&
$[$Ni/Fe$]$&
$[$Ba/Fe$]$\\
& n & $<>$ & rms &    
  n & $<>$ & rms &    
\\
\hline
\multicolumn{16}{c}{Blue HB stars}\\
\hline
~5037 &    &       &      &   &       &      & 0.96 & 0.58 &-0.28 & 0.51 &      &      &      &       &      \\
~6864 &	   &       &      &   &       &      & 0.83 & 0.43 & 0.00 & 0.54 &      &      &      &       &      \\					
~7304 &	   &       &      &   &       &      & 0.68 & 0.07 & 0.31 & 0.51 &      &      &      &       &      \\					
~7432 &	   &       &      &   &       &      & 0.84 & 0.39 & 0.22 & 0.37 &      &      &      &       &      \\					
~8562 &	   &       &      &   &       &      & 1.24 & 0.12 & 0.19 & 0.57 &      &      &      &       &      \\					
\hline
\end{tabular}
\end{scriptsize}
\label{t:tab5}
\end{table*}

\section{Analysis}

The analysis follows procedures similar to those adopted for NGC~2808 
(Paper I) and NGC~1851 (Paper II). Only a few modifications were made.

\subsection{Atmospheric parameters}

For RHB stars, effective temperatures $T_{\rm eff}$\
were derived from the $B-V$ and $V-K$\ colours, using the calibration of Alonso 
et al. (1999, with the erratum of Alonso et al. 2001). The colours were dereddened 
using the $E(B-V)$\ values from the updated on-line version of the Harris (1996) 
catalogue and the $E(V-K)/E(B-V)$\ value from Cardelli et al. (1989). The calibrations 
require input values for the metallicity [A/H]\footnote{We adopt the usual
spectroscopic notation, i.e. [X] = $\log{\rm X_{Star}}-\log{\rm X_\odot}$\ for any
abundance quantity X, and $\log{\epsilon({\rm X})}=\log{\rm N_X/N_H}+12.0$\ for
absolute number density abundances.}. We adopted the values obtained by
Carretta et al. (2009c). The r.m.s. of the differences of the temperatures from
$B-V$\ and $V-K$\ is 56~K for 47~Tuc and 109~K for M~5. The larger scatter found 
for M~5 is expected since stars are fainter and photometric errors larger.

For the blue HB stars in M~5, we started from the $(B-V) - T_{\rm eff}$\ calibration by 
Kurucz\footnote{See kurucz.harvard.edu.}, as for NGC~2808. Infrared colours from 2MASS 
are not reliable for these faint stars. To reduce errors, we averaged temperatures
from $(B-V)$\ colours with those provided from $(V-I)$\ and $(U-B)$. Since $U$\ 
magnitudes are not calibrated, we first obtained the best-fit regression line
between our $U-B$\ colours and $T_{\rm eff}$ from $(B-V)$, and then averaged
the values of $T_{\rm eff}$\ provided by this relation with the other estimates.
We note that the r.m.s. scatter around this relation is 213~K. As in Paper II, we 
assumed errors of 50 and 200~K as representative values for the internal errors in 
the temperatures for RHB and BHB stars, respectively. Systematic errors due to scale 
errors or incorrect parameters for the cluster are likely larger. We discuss below 
their potential impact on our results.

The surface gravities were obtained from the masses, luminosities, and effective temperatures. For 
the masses, we adopted values of 0.648~$M_\odot$\ for the RHB stars of 47~Tuc, and of 0.687 and 
0.573~$M_\odot$\ for stars on the RHB and BHB in M~5 (see Gratton et al. 2010, for a discussion of
values adequate for the different sequences). The BHB mass we consider here was taken from 
the analysis of Gratton et al. (2010).
Masses given there are based on polynomial fits to results from the PISA evolutionary 
models, and might have small errors of up to 0.03~$M_\odot$\ in their values. However,  
such an error in the mass - which would be important if used to compute stellar models - 
has a very minor effect here since even a large difference of 0.05~$M_\odot$\ will only cause a 
difference of only 0.04 dex in the surface gravities, with a negligible impact on the 
abundance analysis. Errors due to other causes (e.g. the assumptions about non-LTE effects)
are much larger. The bolometric corrections
were obtained using calibrations consistent with those used for the effective temperatures (Alonso et
al. 1999 for the RHB stars, and Kurucz for the BHB
stars). The distance moduli were taken from the Harris (1996) catalogue.
Errors in gravities are small. The assumption about masses is likely correct within 10\% (0.04~dex
error in the gravities), while those on the effective temperature and luminosity cause errors in
gravities not larger than $\sim 2$\% for the RHB stars, and $\sim 8$\%
for the BHB stars. The errors in gravities are then not larger than 0.05~dex for the
cool stars and 0.10~dex for the warm ones.

In a preliminary analysis, microturbulence velocities $v_t$\ for RHB stars were obtained by eliminating
any trend of the abundances from Fe\ts{I} line with expected line strength. The r.m.s.
of individual values obtained by this process were 0.18~km$\ts $s$^{-1}$ and 
0.32~km$\ts $s$^{-1}$\ for 47~Tuc and M~5, respectively. However, individual errors 
are responsible for most of this scatter because only a small number of lines spanning 
a limited range in line strength could be measured. For this reason we finally adopted 
the average value of $v_t=1.14$~km$\ts $s$^{-1}$\ for 47 Tuc and 1.42~km$\ts $s$^{-1}$\ for M5 (a
similar approach was adopted in Papers~I and II). For BHB stars (for which no Fe\ts{I} line was 
measured) we adopted values of $v_t$\ as a function of $T_{\rm eff}$\ using a relation 
drawn through data by For \& Sneden (2010). While this last choice is formally different 
from that adopted in Papers~I and II (a constant microturbulence velocity of 
2~km$\ts $s$^{-1}$\ for all BHB stars) in practice the same values were obtained
for stars with the same temperatures.

\subsection{Abundance analysis}

The abundance analysis is very similar to that described in Papers I and II and
is based on equivalent widths. Results are given in Tables~\ref{t:tab3} and
\ref{t:tab4} for 47 Tuc, and in Table~\ref{t:tab5} for M~5.
For Fe lines (only measurable in RHB stars) it is a standard
LTE analysis. For Na and O it is a non-LTE analysis: for RHB stars non-LTE corrections were
taken from Gratton et al. (1999), while for BHB stars they were provided by Takeda (1997) and
Mashonkina et al. (2000). N abundances from N\ts{I} lines in BHB stars also include non-LTE
corrections according to Przybilla \& Butler (2001).

The only significant addition in this paper is the analysis of several CN lines
that could be measured in the spectra of RHB stars of 47~Tuc. Details are given in Sect. 4.3. 

The error analysis is very similar to that given in Paper II. In Table~\ref{t:taberror} we give
the sensitivity of the abundances on the adopted atmospheric parameters as well as
estimates of the overall errors. With the exceptions of N, O, Na and perhaps Ba, the
scatter in the observed abundances agrees very well with these error estimates.

\begin{table*}[htb]
\centering
\caption[]{Sensitivity and error analysis}
\begin{tabular}{lcccccc}
\hline
Parameter &$T_{\rm eff}$&$\log{g}$&$[$A/H$]$&$v_t$& EW & Total \\
\hline
Variation& +100 K & +0.3 dex & +0.2 dex  & +0.5 km/s &+10 m\AA &     \\
Error BHB& 200 K & ~0.1  dex & ~0.05 dex & ~0.5 km/s &~13 m\AA &        \\
Error RHB& ~50 K & ~0.05 dex & ~0.05 dex & ~0.5 km/s &~~4 m\AA &        \\
Error RGB& ~50 K & ~0.05 dex & ~0.05 dex & ~0.15 km/s &~~4 m\AA &        \\
\hline
\multicolumn{7}{c}{Blue HB}\\
\hline
$[$N/Fe$] $ &  0.018 & ~0.049 & -0.001 & -0.057 & 0.092 & 0.138 \\ 
$[$O/Fe$] $ &  0.021 & -0.002 & -0.012 & -0.154 & 0.099 & 0.205 \\ 
$[$Na/Fe$]$ &  0.087 & -0.169 &  0.000 & -0.057 & 0.162 & 0.281 \\ 
$[$Mg/Fe$]$ &  0.000 & ~0.052 & -0.004 & -0.056 & 0.144 & 0.196 \\ 
\hline
\multicolumn{7}{c}{Red HB}\\
\hline
$[$Fe/H$]_I$    & ~0.087 & -0.018 & -0.014 & -0.120 & 0.027 & 0.063 \\ 
$[$Fe/H$]_{II}$ & -0.027 & ~0.126 & ~0.036 & -0.045 & 0.087 & 0.092 \\ 
$[$O/Fe$]$      & -0.208 & ~0.126 & ~0.010 & ~0.065 & 0.087 & 0.138 \\ 
$[$Na/Fe$]$     & ~0.030 & -0.119 & ~0.042 & ~0.047 & 0.075 & 0.081 \\ 
$[$Mg/Fe$]$     & -0.037 & -0.032 & ~0.007 & ~0.048 & 0.150 & 0.152 \\ 
$[$Al/Fe$]$     & -0.043 & ~0.004 & ~0.002 & ~0.105 & 0.106 & 0.113 \\ 
$[$Si/Fe$]$     & -0.051 & ~0.013 & ~0.017 & ~0.067 & 0.075 & 0.082 \\ 
$[$Ca/Fe$]$     & ~0.013 & -0.052 & ~0.018 & ~0.012 & 0.106 & 0.107 \\ 
$[$Ti/Fe$]$     & ~0.039 & ~0.005 & -0.003 & ~0.069 & 0.087 & 0.091 \\ 
$[$V/Fe$]$      & ~0.043 & ~0.006 & ~0.002 & ~0.100 & 0.087 & 0.094 \\ 
$[$Mn/Fe$]$     & ~0.013 & ~0.005 & ~0.000 & ~0.034 & 0.087 & 0.087 \\ 
$[$Ni/Fe$]$     & ~0.018 & ~0.021 & ~0.005 & ~0.051 & 0.087 & 0.098 \\ 
$[$Ba/Fe$]$     & -0.049 & ~0.093 & ~0.077 & -0.311 & 0.150 & 0.180 \\ 
\hline
\end{tabular}
\label{t:taberror}
\end{table*}

\section{Results and discussion}

\subsection{Fe abundances}

The Fe abundances we obtained from this analysis are [Fe/H]=$-0.76\pm 0.01$\ (r.m.s.=0.06~dex)
and $-1.27\pm 0.01$ (r.m.s.=0.06~dex) for 47~Tuc and M~5, respectively (errors here are
simply the standard deviation of the mean values, and do not include systematics).
For comparison, the values listed by Carretta et al. (2009c) are [Fe/H]=$-0.76\pm 0.02$\
and $-1.33\pm 0.02$ for the two clusters. 

As a check of the adopted parameters, we notice that a few Fe\ts{II} lines could be 
measured in RHB spectra. Abundances derived from these lines are in fair agreement with 
those obtained from the Fe\ts{I} lines: on average we obtained [Fe/H]=-$0.87\pm 0.01$\ 
and $-1.35\pm 0.03$ (errors are derived as for Fe~I). The slightly lower abundances obtained from Fe\ts{II} lines might
be explained by various uncertainties in the analysis (e.g. small systematics in the
adopted temperature scale) and by the use of very few
Fe\ts{II} lines. 

Including Papers I and II, we now have an analysis of large samples of RHB 
stars in four GCs made with a homogeneous procedure. Table~\ref{t:FeGCs} lists the 
values we obtain, which compare well to those for RGB stars from Carretta 
et al. (2009c).

\begin{table}[htb]
\centering
\caption[]{Comparison between Fe abundances from RHB and RGB stars}
\begin{tabular}{lcccccc}
\hline
NGC & N &\multicolumn{4}{c}{$[$Fe/H$]_{\rm RHB}$} & $[$Fe/H$]_{\rm RGB}$ \\
    & RHB & $[$Fe/H$]{\sc I}$ & r.m.s.& $[$Fe/H$]{\sc II}$ & r.m.s. & \\
\hline
~104 & 110 & -0.76 & 0.06 & -0.88 & 0.09 & $-0.76\pm 0.02$ \\
1851 &  54 & -1.18 & 0.06 & -1.20 & 0.10 & $-1.18\pm 0.08$ \\
2808 &  36 & -1.18 & 0.07 & -1.13 & 0.17 & $-1.18\pm 0.04$ \\
5904 &  30 & -1.27 & 0.06 & -1.35 & 0.14 & $-1.33\pm 0.02$ \\
\hline
\end{tabular}
\label{t:FeGCs}
\end{table}

\begin{center}
\begin{figure}
\includegraphics[width=8.8cm]{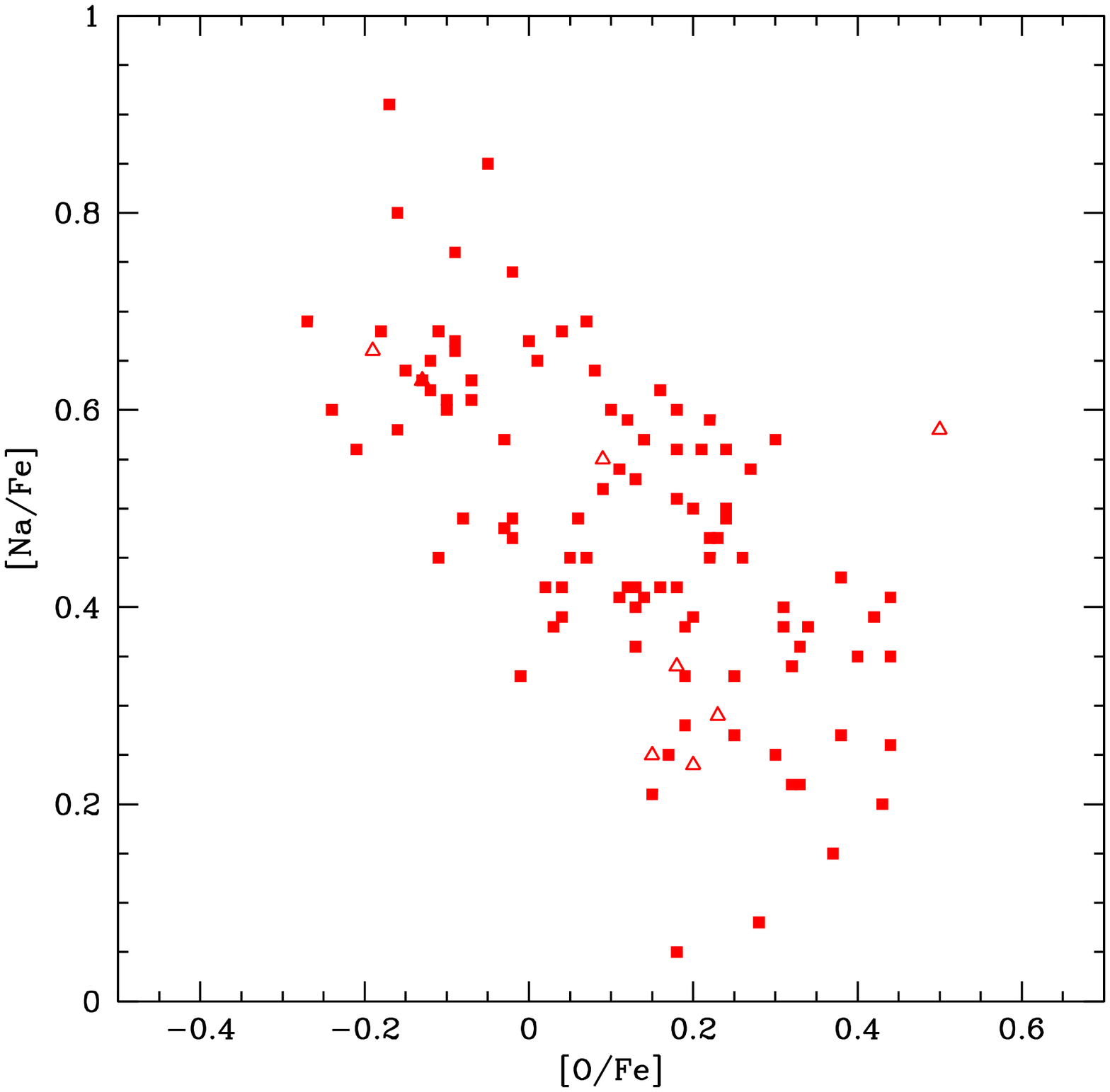}
\includegraphics[width=8.8cm]{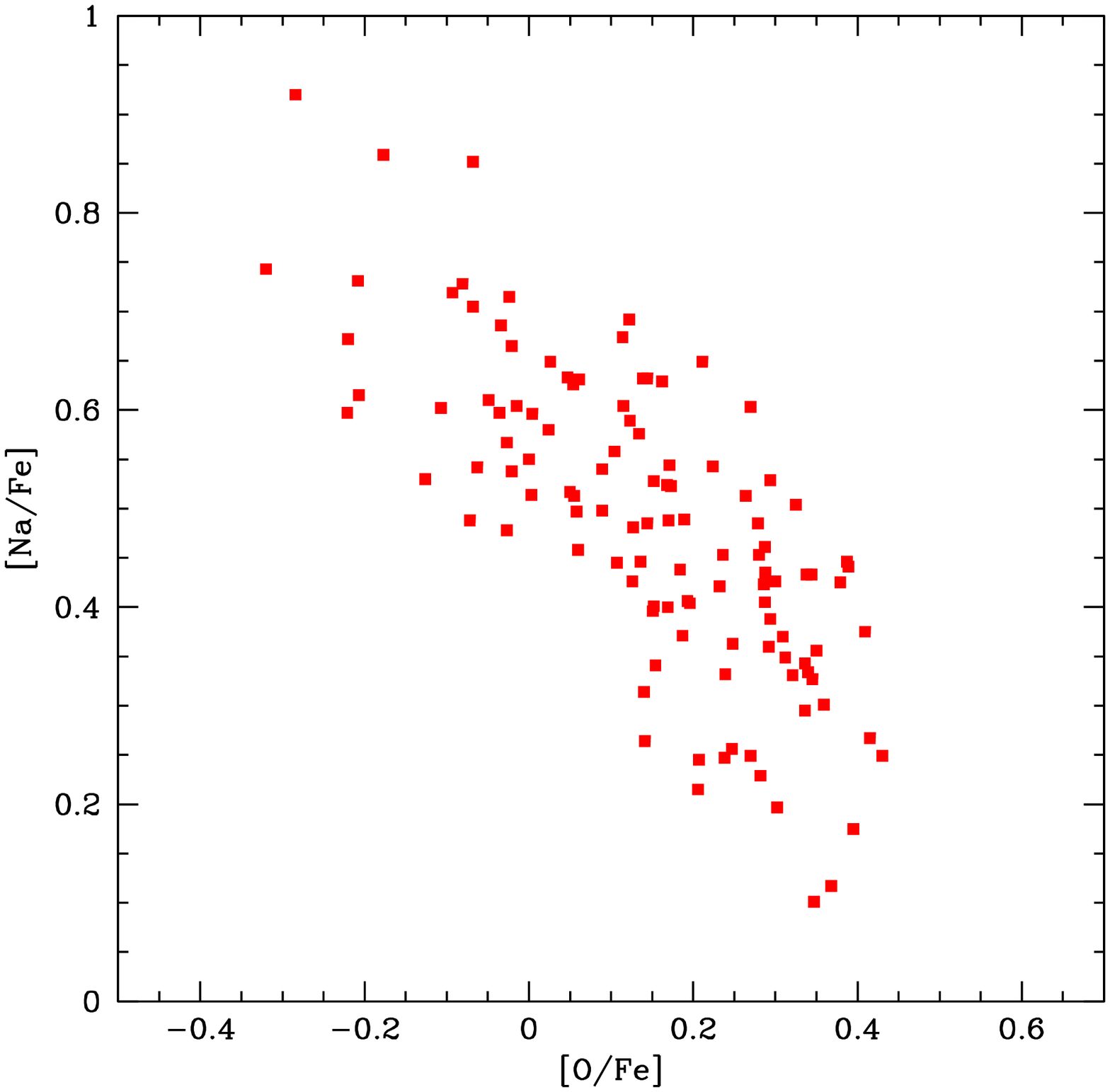}
\caption{Upper panel: Na-O anticorrelation for HB stars in 47~Tuc; filled squares are the faint stars
group, open triangles are the bright stars group. Lower panel: the same for RGB stars 
(from Carretta et al. (2009a) }
\label{f:fig4}
\end{figure}
\end{center}

\begin{center}
\begin{figure}
\includegraphics[width=8.8cm]{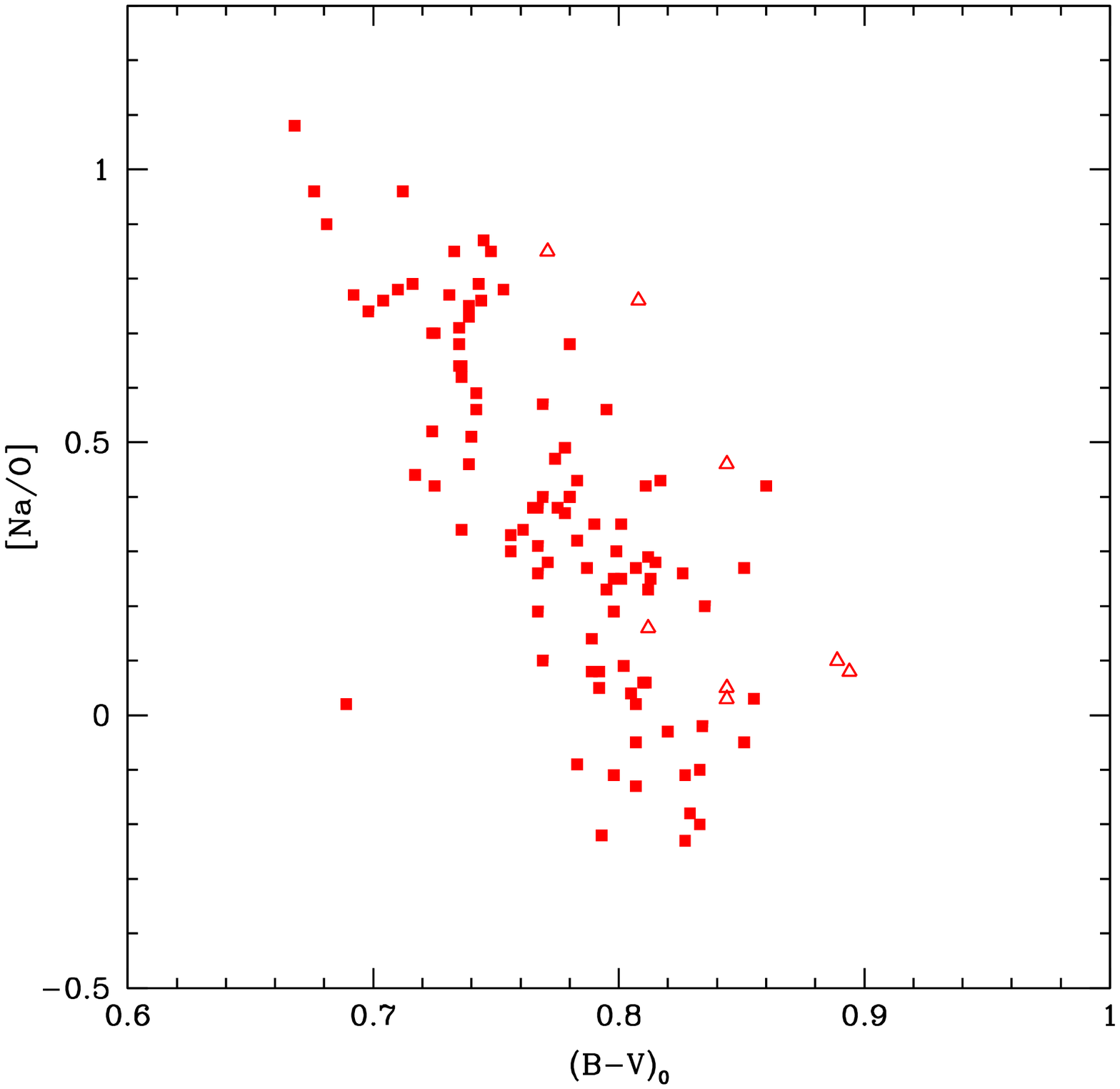}
\caption{Run of the [Na/O] ratio with $B-V$\ colour along the HB of 47~Tuc. Symbols are as in 
Figure~\ref{f:fig4}. }
\label{f:fig5}
\end{figure}
\end{center}

\begin{center}
\begin{figure}
\includegraphics[width=8.8cm]{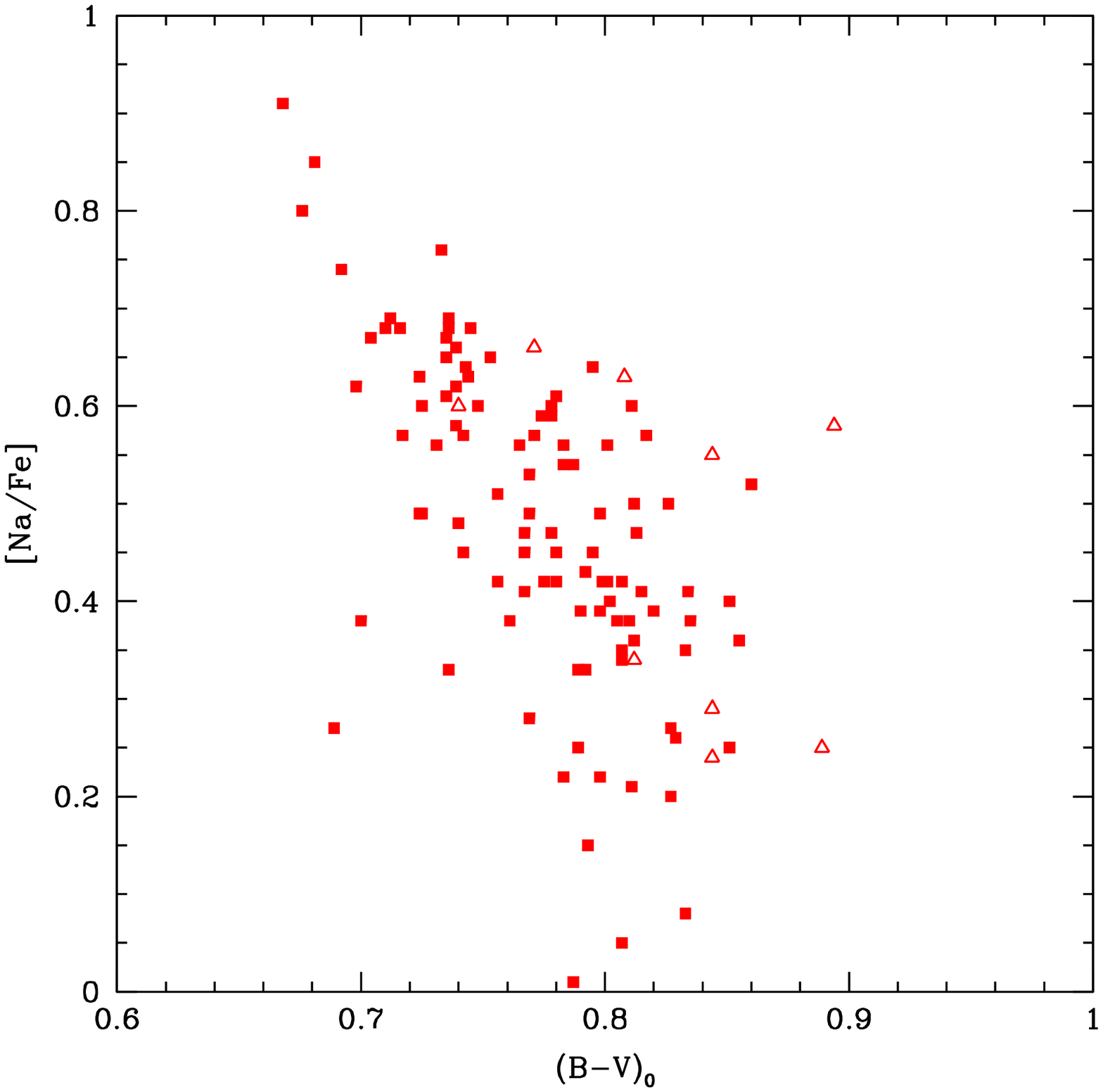}
\includegraphics[width=8.8cm]{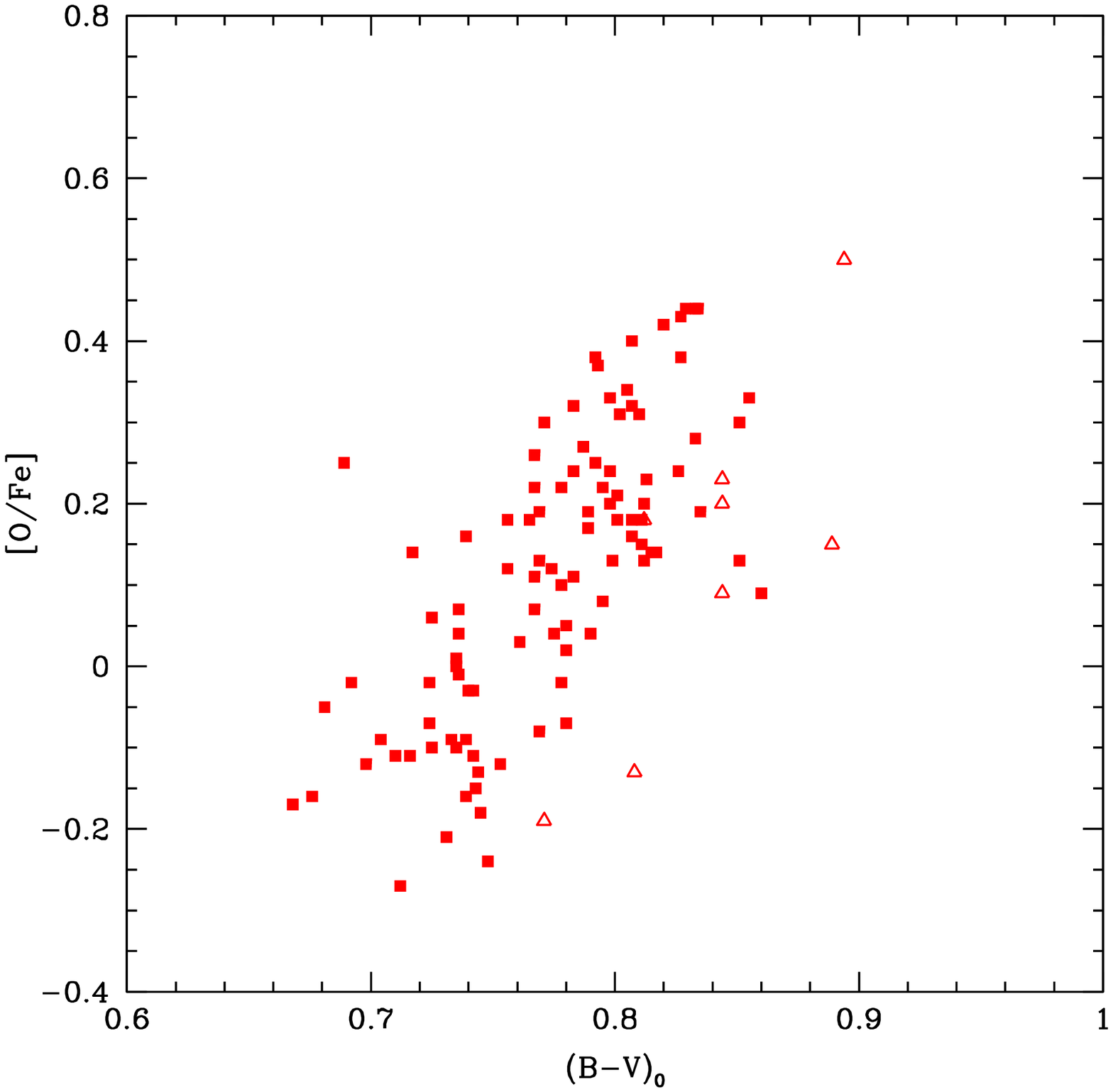}
\caption{Upper panel: run of the [Na/Fe] ratio with $B-V$\ colour along the HB of 47~Tuc;
lower panel: the same for the [O/Fe] ratio. Symbols are as in 
Figure~\ref{f:fig4}. }
\label{f:fig6}
\end{figure}
\end{center}

\begin{table*}[htb]
\centering
\caption[]{Average abundances}
\begin{tabular}{lcccccccccccc}
\hline
Element &\multicolumn{2}{c}{47~Tuc} &\multicolumn{4}{c}{M~5}&\multicolumn{4}{c}{NGC~1851}&\multicolumn{2}{c}{NGC~2808}\\
   &\multicolumn{2}{c}{RHB} &\multicolumn{2}{c}{RHB}&\multicolumn{2}{c}{BHB}&\multicolumn{2}{c}{RHB}&\multicolumn{2}{c}{BHB}&\multicolumn{2}{c}{RHB}\\
        & $<>$ & rms & $<>$ & rms & $<>$ & rms & $<>$ & rms & $<>$ & rms & $<>$ & rms \\
\hline
$[$N/Fe$]$  &~1.62 & 0.21 &      &      & 0.68 & 0.27 &      &      & 1.06 & 0.16 &      &      \\
$[$O/Fe$]$  &~0.11 & 0.18 &~0.53 & 0.13 & 0.35 & 0.19 & 0.37 & 0.12 & 0.01 & 0.40 & 0.44 & 0.16 \\
$[$Na/Fe$]$ &~0.48 & 0.17 &~0.14 & 0.12 & 0.20 & 0.26 & 0.11 & 0.25 & 0.64 & 0.35 & 0.14 & 0.10 \\
$[$Mg/Fe$]$ &~0.29 & 0.09 &~0.20 & 0.15 & 0.30 & 0.21 & 0.43 & 0.06 & 0.42 & 0.23 & 0.22 & 0.14 \\
$[$Al/Fe$]$ & 0.16 & 0.10 &      &      &      &      & 0.25 & 0.15 &      &      &      &      \\
$[$Si/Fe$]$ &~0.23 & 0.06 &~0.29 & 0.11 &      &      & 0.24 & 0.11 &      &      & 0.38 & 0.16 \\
$[$Ca/Fe$]$ &~0.45 & 0.10 &~0.54 & 0.13 &      &      & 0.47 & 0.15 &      &      & 0.43 & 0.16 \\
$[$Ti/Fe$]$ &~0.17 & 0.07 &      &      &      &      &      &      &      &      &      &      \\
$[$V/Fe$]$  &-0.05 & 0.08 &      &      &      &      &      &      &      &      &      &      \\
$[$Mn/Fe$]$ &-0.42 & 0.06 &-0.61 & 0.17 &      &      &      &      &      &      &      &      \\
$[$Ni/Fe$]$ &-0.01 & 0.06 &-0.38 & 0.07 &      &      &      &      &      &      &      &      \\
$[$Ba/Fe$]$ &~0.38 & 0.16 &~0.00 & 0.22 &      &      & 0.33 & 0.39 &      &      & 0.30 & 0.19 \\
\hline
\end{tabular}
\label{t:average}
\end{table*}

Table~\ref{t:average} lists the average abundances for
47~Tuc, M~5 (this paper), and NGC~1851 (Paper II). The Al abundance in NGC~1851
RHB stars obtained in Paper II was corrected upward by 0.2 dex because we realized it was
obtained by assuming an inconsistent, too high, solar Al abundance. In most cases, we obtained for all
these clusters the classical pattern observed in metal-poor stars: overabundances
of $\alpha-$elements (Mg, Si, Ca, Ti) and underabundance of Mn. In the rest of this
section we examine evidence for the light elements (He, N, O, Na, Mg, and Al)
and for the only $n-$capture element we observed (Ba).

\subsection{Na-O anticorrelation}

\subsubsection{47~Tuc}

Figure~\ref{f:fig4} compares the Na-O anticorrelation obtained from RHB
stars in 47~Tuc with that obtained for RGB by Carretta et al. (2009a).
For this comparison, we checked that our selection of HB
stars in 47~Tuc does not miss any significant group of stars. However, the
presence of selection effects in magnitude (see Section 2) has a significant 
effect since the fraction of stars for which we acquire spectra is higher
among the bluer part of the HB than the red one. As we will see below,
there is a correlation between colour of the stars and their [Na/O] ratio.
The selection effect considered above accordingly produces a higher percentage of
Na-rich and O-poor stars in our sample than are really present along the HB
of 47~Tuc. The fraction of stars with [Na/O]$>0.5$\ in our sample is 
$\sim 50$\% too high because of this effect. However, the impact of this effect 
on the overall distribution of stars along the Na-O anticorrelation is small
and the distribution obtained for RHB stars still looks similar 
to that obtained for RGB stars. The interquartile of the [Na/O] ratio for RHB is 
0.48~dex, to be compared with a value of 0.47~dex obtained for RGB stars (see Carretta 
et al. 2010). We recall that the interquartile is the range in the quantity - in
this case the [Na/O] ratio - including 50\% of the distribution.

In Figure~\ref{f:fig5} we plot the run of the [Na/O] ratio against
$B-V$\ colour along the HB of 47~Tuc. There is a very good correlation
with increasing [Na/O] ratios for decreasing $B-V$\ colour (that is,
increasing $T_{\rm eff}$). If we only consider stars of the faint group, the
correlation coefficient is $r=0.74$ over 99 stars, which is more than 
a 10~$\sigma$\ effect. This trend is caused by both a decrease in [O/Fe]
and an increase in [Na/Fe] with colors becoming bluer (see Figure~\ref{f:fig6}).

This trend is much stronger than can be explained by observational and
systematic errors, and should therefore be true. It implies that the location
of a star along the bright part of the HB of 47~Tuc can be predicted with quite a good
accuracy if its [Na/O] ratio is known. It recalls the similar
results found by Marino et al. (2011) for M~4 and by us (Paper I) for NGC~2808.
It can be understood if the variation of the [Na/O] ratio is related to that
of He, and the mass of the stars in evolution along the HB is in turn mainly
related to their He content. In Section 2 we estimated that the masses of HB 
stars of 47~Tuc range from $\sim 0.60$\ up to $\sim 0.71$~$M_\odot$ (peak to 
valley); this corresponds to a small range in He abundances of 
$\Delta Y<0.03$ (Di Criscienzo et al. 2010; Nataf et al. 2011; Milone et al. 
2012). The good predictive power of the [Na/O] ratio on the location of a star along the
HB of 47~Tuc argues against large random effects (that is, effects that
are independent of the Na-O anticorrelation) in the mass loss, at
least for this cluster.

We may try to provide an upper limit to such random effects by considering 
the spread in $B-V$\ colours around the best-fitting relation with [Na/O].
This is 0.029~mag. However, part of this spread is caused by observational errors
in the colours (that we may roughly estimate to be $\sim 0.01$~mag), in the [Na/O]
values ($\sim 0.1$~dex, which given the slope of the relation also implies
an effect of $\sim 0.01$~mag in $B-V$), and evolution of stars off the ZAHB, which
does not occur at exactly constant colours. 


\begin{center}
\begin{figure}
\includegraphics[width=8.8cm]{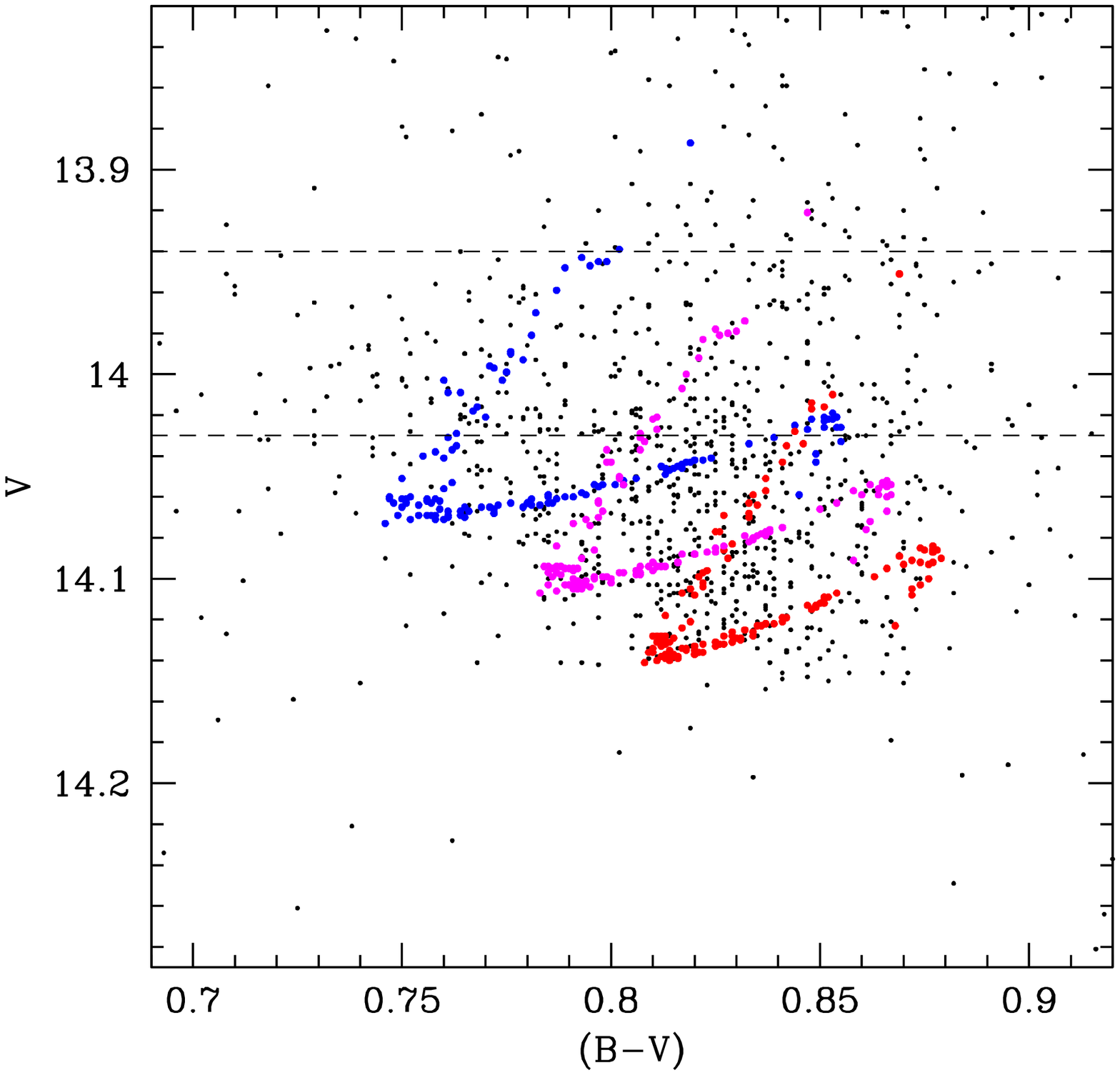}
\caption{HB stars of 47 Tuc in the $V-(B-V)$\ CMD (black dots). Dashed horizontal
lines mark the upper and lower magnitude limit of our faint sample. Some synthetic 
HB models computed by adopting three distinct initial He abundances
(Y=0.26, blue dots, Y=0.27, magenta dots; and Y=0.28, red dots) are also shown. See
text for details.}
\label{f:fighb1}
\end{figure}
\end{center}

Additional insight is gained by comparing these data with synthetic HBs. These last
were estimated using the same procedure as adopted for NGC~1851 (Gratton
et al. 2012b) and were computed with a range of different He abundances. For all 
He abundances, the same average amount of mass lost during the RGB has been adopted.
The value adopted is ${\rm 0.224M_\odot}$\ with a negligible dispersion around it.
Note that this is a constant mass offset with respect to that predicted by
models of same age but different He abundances at the tip of the RGB, and does not 
assume strictly the same mass loss law. To compare observations with theoretical 
predictions, we adopted a reddening $E(B-V)=0.04$ and a distance modulus 
${\rm (m-M)_V}=13.43$. These values are a small adaptation of those listed in the 
Harris (1996) catalogue to provide a best match to observations.

In this comparison, we first notice that 
the HB of 47 Tuc has a roughly triangular shape (see Figure~\ref{f:fighb1}). This matches
the predictions well if we assume a small spread in He: indeed, He-rich stars are not
only brighter, but also describe wider loops while evolving than He-poor stars.
As found previously by Di Criscienzo et al. (2010), a small spread of $\Delta Y$=0.02-0.03
can well explain the morphology of the HB of 47 Tuc. We now focus on our faint
sample, which spans the whole range in colour of the HB of 47 Tuc, but consists of stars 
with $13.94<V<14.03$. This is a narrow strip at the bright extreme of the most populated
region of the HB of 47 Tuc. Within this strip, a correlation between He abundance and
colour might be expected if there is no significant fraction of stars with $Y\geq 0.28$.
HB stars with He abundances lower than this value cross the observed strip only
during their late evolution at magnitudes brighter than the ZAHB, at a colour that
depends on their He abundance. Stars with $Y>0.28$ would rather spend most of their HB 
life within the observed magnitude strip, making a wide loop in colour: many of them would be
quite red, and since they would also likely be Na-rich and O-poor, the correlation
between $B-V$\ colour and Na/O should be much weaker than observed.

A more detailed comparison between observations and theoretical predictions
is hampered by several - usually neglected - possible sources of uncertainty that may become important
at this level. A short list would include the treatment of superadiabatic convection and the
color-${\rm T_{eff}}$ relations used for transferring the models to the observational planes. 
However, three conclusions are likely robust: (i) There
is a spread in He abundances within 47 Tuc; (ii) this spread is narrow ($0.26<Y<0.28$); and
(iii) there is only very limited room for star-to-star variations in the total mass lost
when ascending the RGB.

This last result apparently contrasts with the finding of strong star-to-star variations
in the mass loss rates found by some authors for red giants in 47 Tuc by modelling the 
H$\alpha$ profiles (see e.g. McDonald \& van Loon 2007). While we underline that ours is an 
empirical result based on evolutionary models, we notice that the mass loss rate measures
are instantaneous values that may differ substantially from the values integrated over
the whole RGB phase, since it is very likely that mass loss is episodic (see e.g. Origlia et al.
2007). We also note that our result does not exclude that a small minority of stars in 47 Tuc
have very different mass loss histories due e.g. to evolution within binary systems (see e.g.
Mohler et al. 2000 and Knigge et al. 2008).

\begin{center}
\begin{figure}
\includegraphics[width=8.8cm]{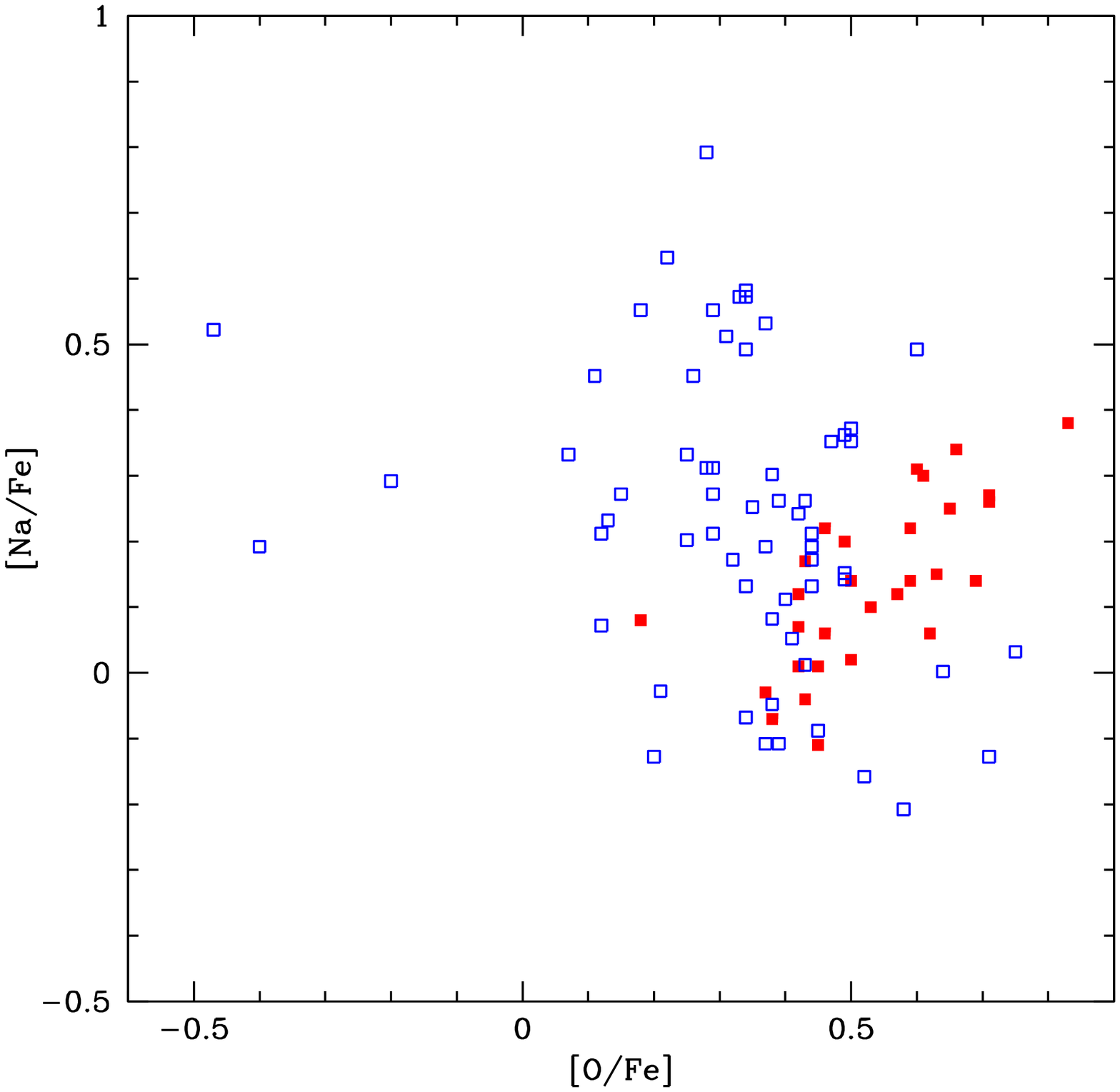}
\includegraphics[width=8.8cm]{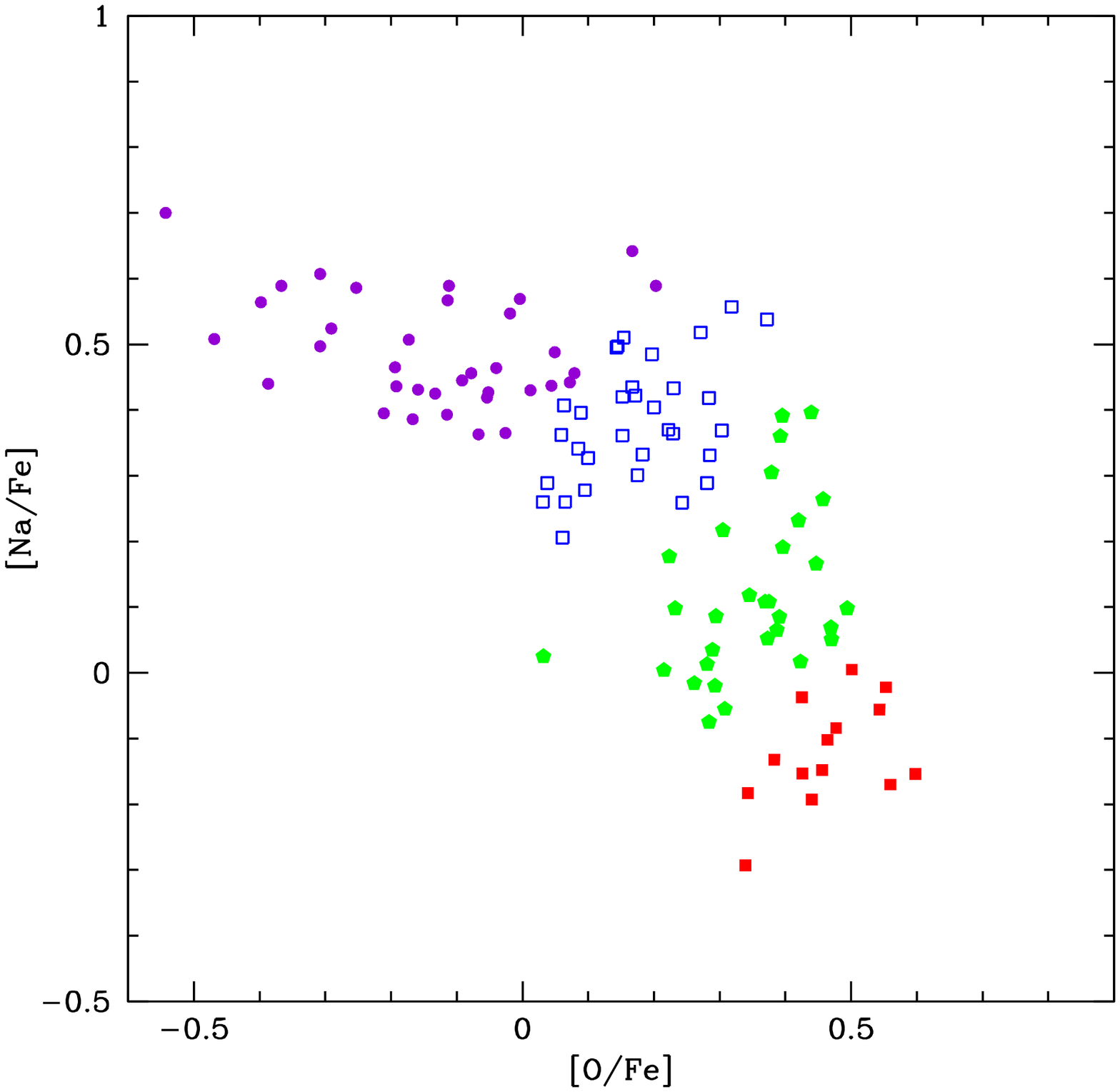}
\caption{Upper panel: Na-O anticorrelation for HB stars in M~5; filled red squares are RHB stars, blue
open squares are BHB stars. Lower panel: the same for RGB stars (from Carretta et al. (2009a). In this 
panel, we plotted with different symbols stars that might be expected to have descendants
on the VBHB (stars bluer than the Grundahl jump: filled violet circles), BHB (open blue squares), RR 
Lyrae (filled green pentagons), and RHB (filled red squares) (see text). }
\label{f:fig6b}
\end{figure}
\end{center}

\begin{center}
\begin{figure}
\includegraphics[width=8.8cm]{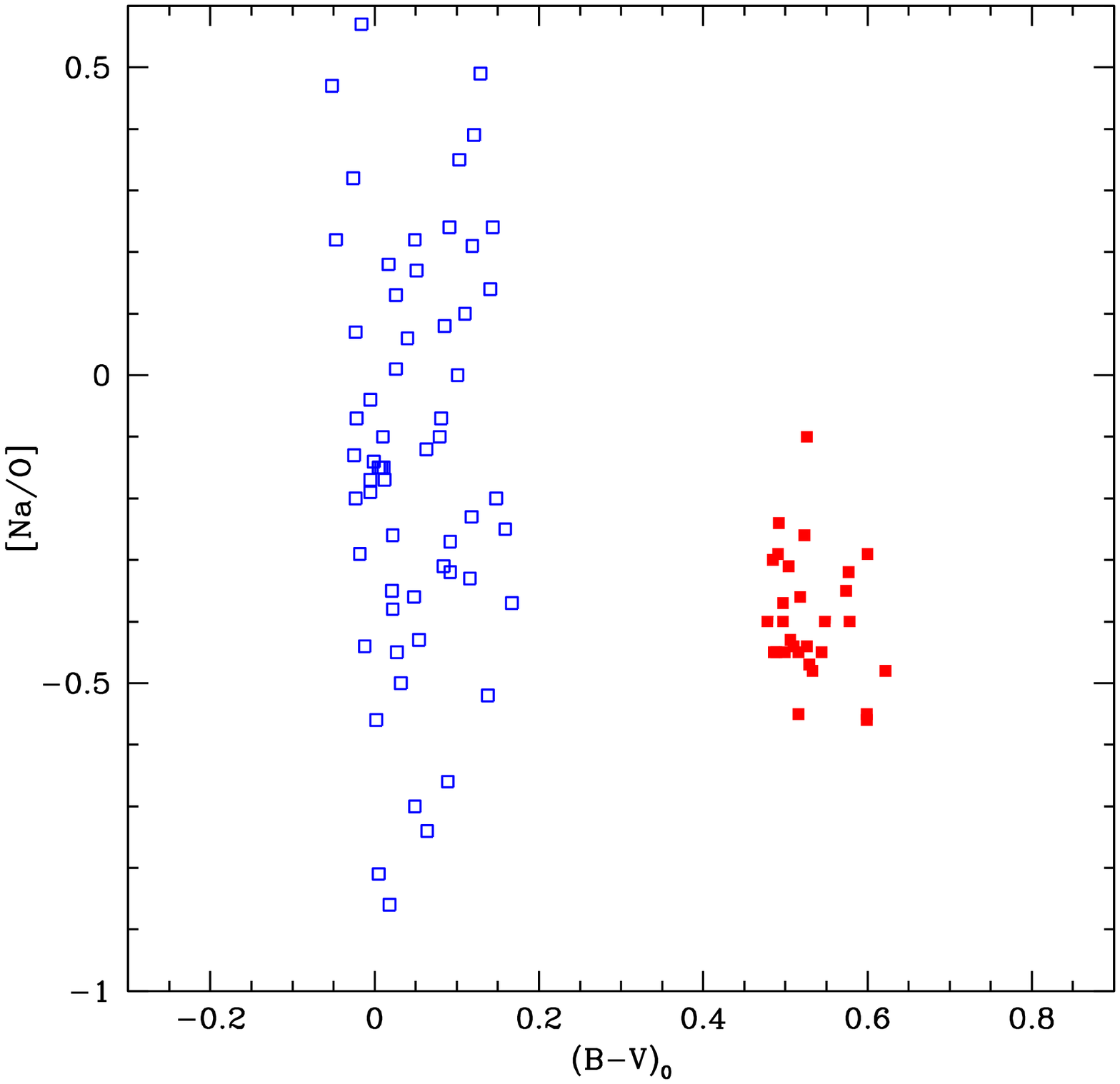}
\caption{Run of the [Na/O] ratio with $B-V$\ colour along the HB of M~5. 
Symbols are as in Figure~\ref{f:fig6b}.}
\label{f:fig7}
\end{figure}
\end{center}

\begin{center}
\begin{figure}
\includegraphics[width=8.8cm]{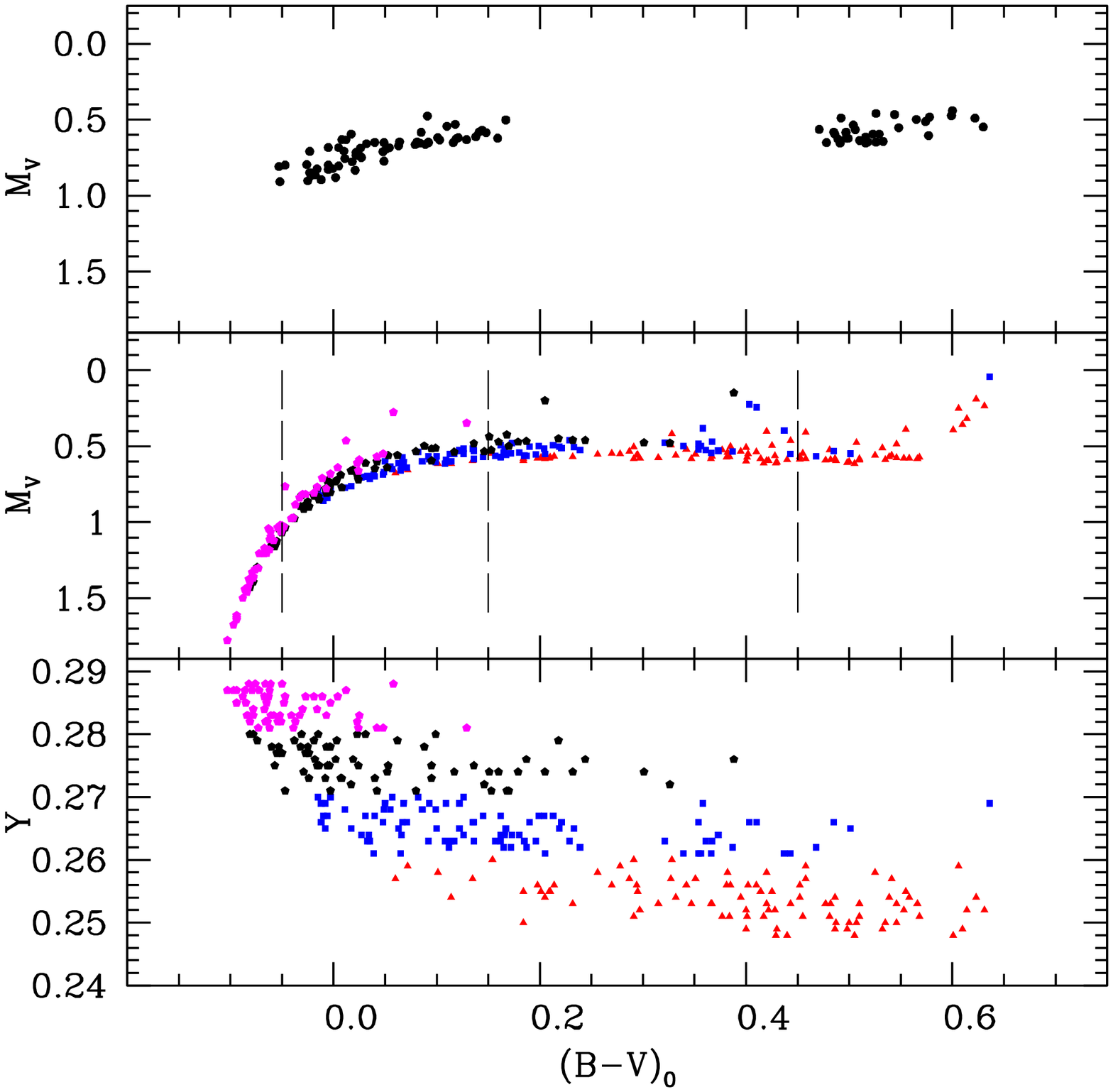}
\caption{Upper panel: observed HB stars of M5 in the $M_V-(B-V)_0$\ CMD. Middle panel: synthetic HB
computed with parameters appropriate to M~5, and a uniform distribution of helium
abundances in the range $0.248<Y<0.288$. Stellar masses assume a Gaussian distribution
with an r.m.s. of $0.03~M_\odot$ around a value that is the same for all He contents.
When transforming a theoretical CMD into an observational one, we assume a reddening of 
$E(B-V)=0.03$\ and a distance modulus of $(m-M)_V=14.46$. Lower panel: run of He abundances
for the simulated stars with $B-V$\ colour. }
\label{f:fighb2}
\end{figure}
\end{center}

\subsubsection{M~5}

Figure~\ref{f:fig6b} compares the Na-O anticorrelation for stars on the HB of M~5
with those on the RGB (these last are taken from Carretta et al. 2009a). To understand
this comparison we recall that at variance with 47~Tuc, M~5 has a very 
broad HB and our sample has considerable selection effects. We may divide the HB of M~5
into four sections:
\begin{itemize}
\item Very blue HB stars (VBHB, here stars with $M_V<0.9$): these stars are warmer than 
the Grundahl jump. We estimated that they represent 33\% of the total HB stars from 
counts on the HST data by Piotto et al. (2002), analysed as in Gratton et al. (2010). 
Their atmospheres are in radiative equilibrium and the chemical abundances are very 
different from the original ones due to sedimentation and radiative levitation effects 
(see Behr et al. 1999). For this reason we did not include these stars in our sample.
\item BHB stars, here stars redder than VBHB stars, but bluer than the instability
strip, that is with $(B-V)_0<0.15$. Following the same procedure as described above,
we estimate that these stars represent 28\% of the M~5 HB population. We have 61 such 
stars in our sample.
\item RR Lyrae variables. M~5 has a very rich population of these variables, representing
some 26\% of the total HB population. We did not observe them because they should 
be observed close to minimum to obtain reliable results, which is not practical for 
multi-object observations in service mode.
\item RHB stars, that is, stars redder than the instability strip. They represent about 13\% 
of the M~5 HB stars. We observed 30 stars in this group; however, a significant fraction
of them (12 out of 30) are distinctly brighter than the others, and are
probably evolved objects on their way towards the asymptotic giant branch (AGB),
although they may be also massive objects resulting from the evolution of binary
systems. Incidentally, the ratio of BHB/RHB among the stars we analysed is close to the actual ratio of BHB/RHB
stars in M~5.
\end{itemize}

Also for comparison, we recall that Carretta et al. (2009a) classified 27\% of the
RGB stars as belonging to the primordial population, 66\% to the intermediate
population, and 7\% to the extreme population for this cluster. If the colours 
of the stars along the HB are mainly influenced by their He content and the [Na/O] 
ratio, the Na-O anticorrelation is not completely sampled by the HB stars considered 
in this paper. We expect to find very few extremely O-poor stars, because they likely 
spend most of their HB life as VBHB stars. However, a few of them might be present because 
at the late stages of their HB evolution, stars originally on the VBHB cross the BHB 
region. We likely also missed several moderately Na-rich/O-poor stars that probably are on the 
instability strip. 

To clarify these rough expectations (which neglect the effect of
evolution off the ZAHB, which is significant), we marked with different symbols stars with different
values of the [Na/O] ratio in the right panel of Figure~\ref{f:fig6b}; the descendants
of these different groups of stars are expected to mainly populate the VBHB, BHB, 
the instability strip, and the RHB according to their [Na/O] ratio. 
Once this is taken into account,
we find that, as expected, RHB stars have high O and low Na abundances (see also
Figure~\ref{f:fig7}, where we display the run of the [Na/O] ratio with $B-V$\ 
colours), and that we missed the extreme O-poor stars almost completely, likely 
because they are on the VBHB. On the other hand, while we indeed found a Na-O
anticorrelation among BHB stars, the [Na/O] values of most stars 
on the BHB are similar to those of stars on the RHB (they can therefore be 
interpreted as objects belonging to the primordial population), and the scatter of 
[Na/O] values for these stars is large. Part of these results may be attributed to 
observational errors (which are quite large for BHB stars) and to evolution off the 
ZAHB. However, at variance with the case of 47 Tuc, presumably some additional scatter 
in mass loss should also be included to explain our data for M~5. This is suggested by
the comparison with a synthetic HB shown in Figure~\ref{f:fighb2}. The theoretical
computations were made in the same way as for 47~Tuc; however, in this case we assumed
that helium is uniformly distributed in the range $0.248<Y<0.288$, but we also have
to assume that mass loss has a Gaussian distribution with an r.m.s. of $0.03~M_\odot$\
to have a reasonable number of He-poor stars on the BHB. Note that this synthetic HB
does not reproduce the distribution of stars along the HB of M~5 in detail since
stars at the blue extreme of the HB are under-represented. According to the definitions given
above, the synthetic HB contains about 7\% VBHB against an observed population of 33\%.
This indicates that there are more very He-rich stars than assumed in this simulation. 
This is not relevant in the present discussion, which focuses on a few qualitative features 
that are required to explain the observations, which exclude VBHB stars, however. The lower panel of
Figure~\ref{f:fighb2} displays the run of He with colours for stars in this synthetic
HB. The distribution of stars in this plot matches the observed run with
colours of the [Na/O] and [N/Fe] abundance ratios quite well, if the quite large
observational errors in the [Na/O] values for BHB stars are taken into account.

We conclude that the observations of M~5 on the whole do not contradict the scheme
where variations in the He abundances might contribute to explain the HB of this cluster; 
however, in this case we should also assume the presence of an additional source of 
substantial scatter in the mass loss rates of stars climbing up the RGB. This
is reminiscent of the small range in He abundances along the HB of M~3, as discussed by
Catelan et al. (2009); a similar argument was already presented for the same M~5 by
Crocker et al. (1988). On the other hand, we should recall that the relation between
variations in He and Na and O abundances may be not simple (see e.g. Marcolini et al.
2009).

\begin{center}
\begin{figure}
\includegraphics[width=8.8cm]{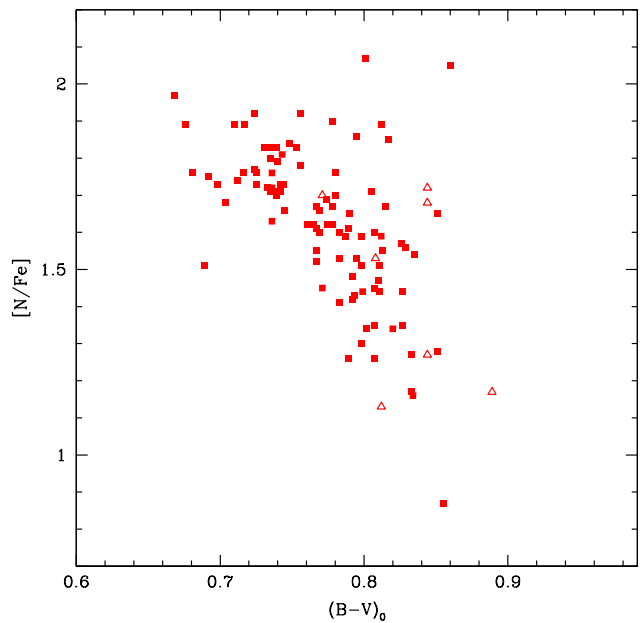}
\caption{Run of the [N/Fe] ratio with $B-V$\ colour along the RHB of 47~Tuc. Symbols are as in Figure 4.}
\label{f:fig8}
\end{figure}
\end{center}

\begin{center}
\begin{figure}
\includegraphics[width=8.8cm]{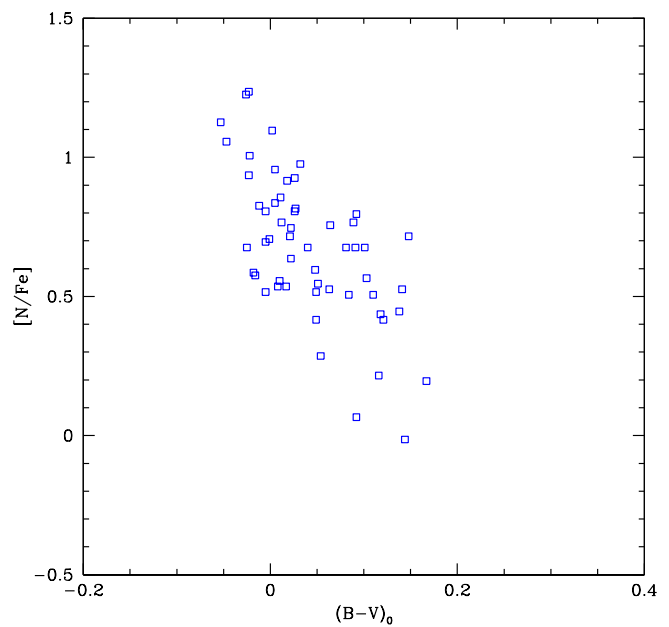}
\caption{Run of the [N/Fe] ratio with $B-V$\ colour along the BHB of M~5. Symbols are as in Figure~\ref{f:fig6b}.}
\label{f:fig9}
\end{figure}
\end{center}

\subsection{Nitrogen}

We were able to derive a hint about the N abundance for stars on the RHB of 47~Tuc, using a
number of CN lines. The actual N abundance depends on the value we assume for the
C abundance, for which we have no diagnostic. We assumed [C/Fe]=-0.6, which is
a typical value for field HB stars (Gratton et al. 2000). Figure~\ref{f:fig8}
displays the run of the [N/Fe] abundances with $(B-V)$ colour in 47~Tuc: there is a very
tight anti-correlation. Since we expect that Na-rich/O-poor stars be also N-rich (O
being transformed into N), this result strongly supports that obtained for the 
[Na/O] ratio.

CN lines are too weak to be measurable in our spectra of RHB stars of M~5. On
the other hand, the N\ts{I} line at 8216~\AA\ could be measured on the spectra of BHB stars (as we
did for BHB stars in NGC~1851, see Paper II). Figure~\ref{f:fig9}
displays the run of these [N/Fe] abundances with $(B-V)$ colour. In this
case we also obtain an anticorrelation, mainly driven by the high N abundances
we obtain for the warmer stars. Again, this supports the result we obtained
from the [Na/O] ratio.

\begin{center}
\begin{figure}
\includegraphics[width=8.8cm]{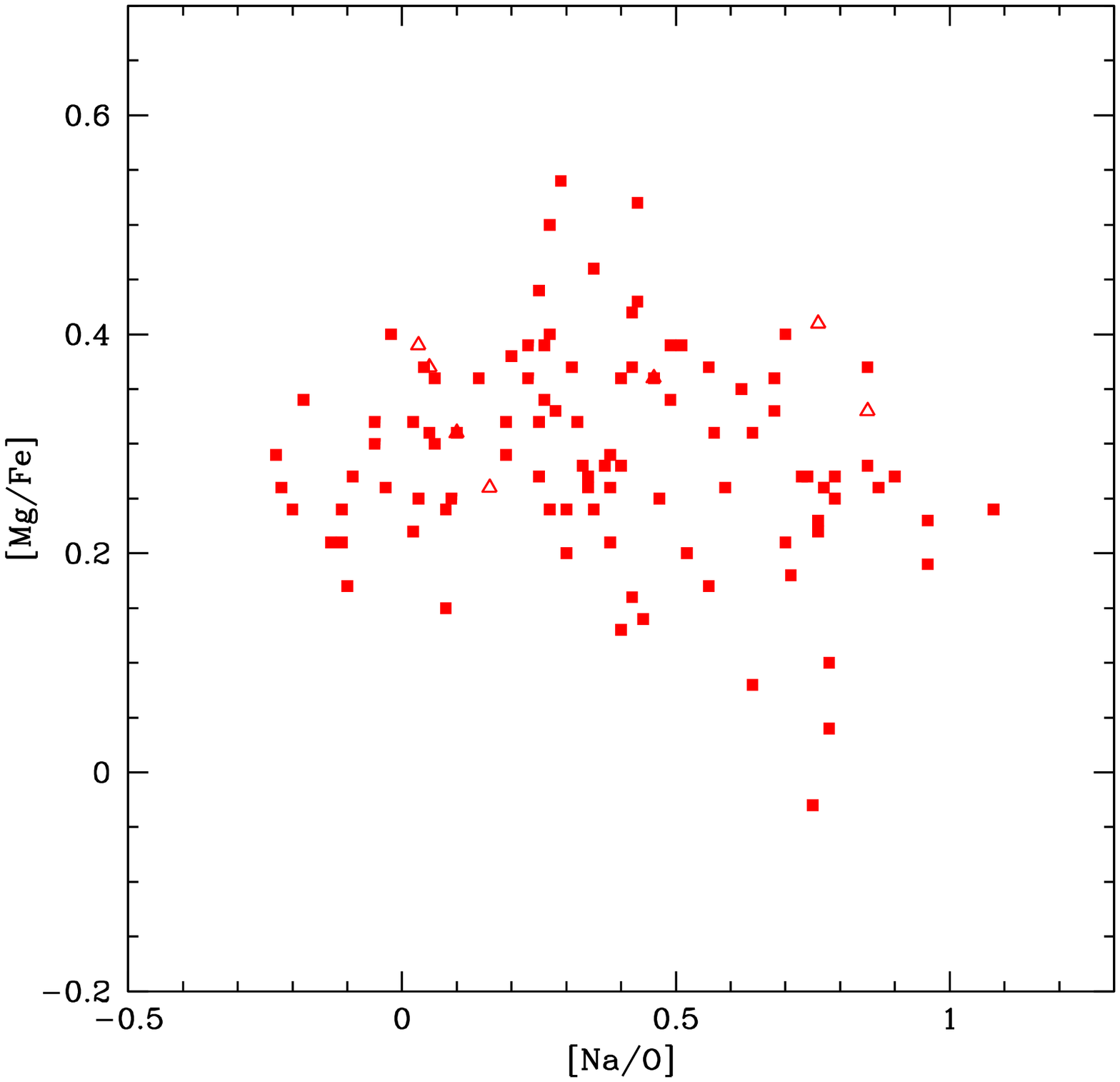}
\includegraphics[width=8.8cm]{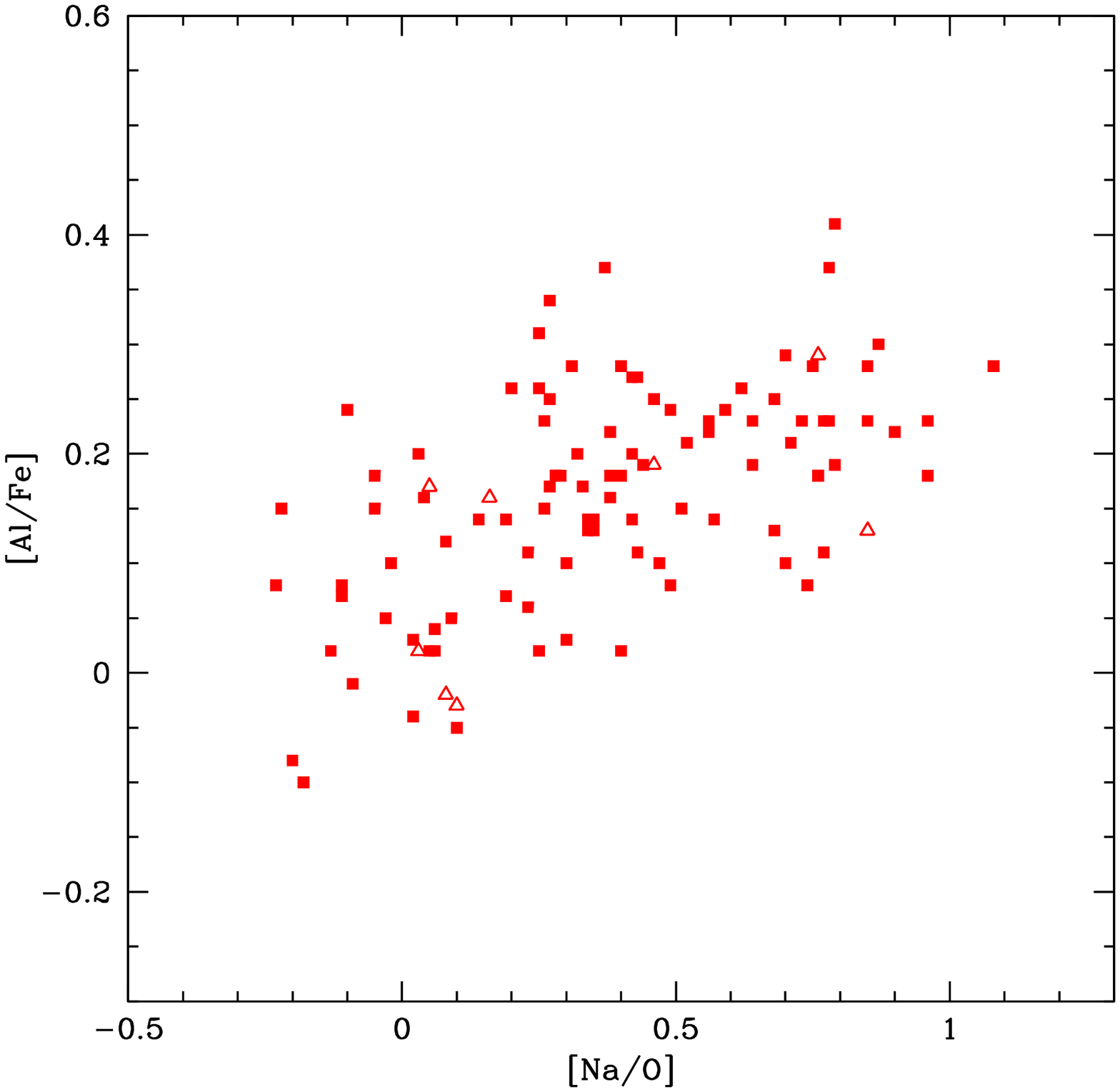}
\caption{Upper panel: run of the [Mg/Fe] ratio with [Na/O] ratio for stars in 47~Tuc.
Lower panel: the same for the [Al/Fe] ratio. Symbols are as in Figure 4.}
\label{f:fig10}
\end{figure}
\end{center}

\begin{center}
\begin{figure}
\includegraphics[width=8.8cm]{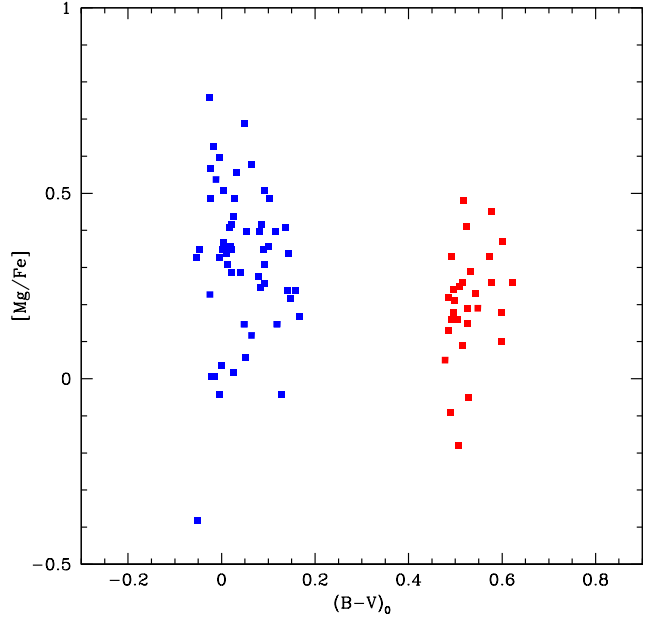}
\caption{Run of the [Mg/Fe] ratio with $(B-V)$\ colour for stars in M~5.
Symbols are as in Figure~\ref{f:fig6b}. }
\label{f:fig11}
\end{figure}
\end{center}

\subsection{Mg and Al}

Both Mg (from the Mg\ts{I} line at 8213~\AA) and Al (from the doublet at 7835-36~\AA) 
abundances were obtained for RHB stars in 47~Tuc. Figure~\ref{f:fig10} displays the 
run of [Mg/Fe] and [Al/Fe] with [Na/O] for stars along the HB of 47~Tuc. We found a 
clear correlation between [Al/Fe] and [Na/O], while there is no convincing trend for 
[Mg/Fe]. This result is not surprising, given that the range of [Al/Fe] is quite 
limited ($\sim 0.4$)~dex, and significant Mg depletion is only expected for [Al/Fe]$>1$.

Al lines are too weak to be detected in RHB stars of M~5. The run of [Mg/Fe] with 
$(B-V)$\ colour for stars in this clusters is shown in Figure~\ref{f:fig11}.
Mg abundances were derived from the Mg\ts{I} line at 8213~\AA\ in RHB stars, and from
the Mg\ts{II} lines at 7877-96~\AA\ in BHB stars. Results for most of the stars are
compatible with a single Mg abundance within the errors. Indication for a
high Mg depletion is only obtained for one of the hottest BHB star (\#25489). 

\begin{center}
\begin{figure}
\includegraphics[width=8.8cm]{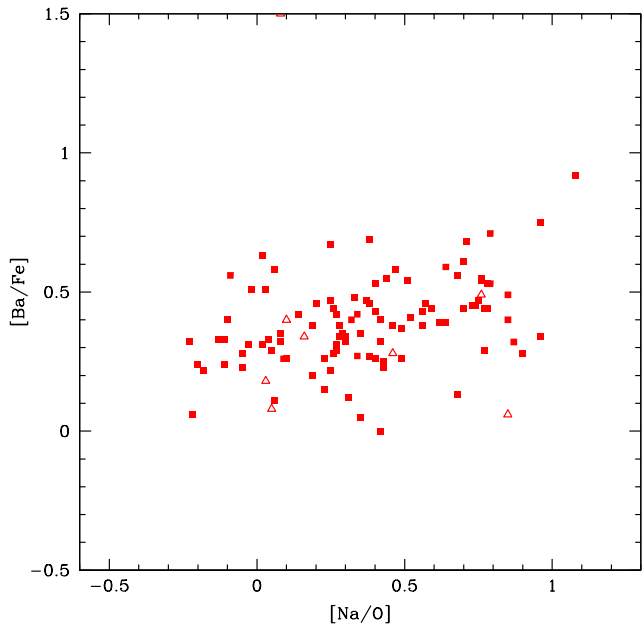}
\caption{Run of the [Ba/Fe] ratio with [Na/O] ratio for stars in 47~Tuc. 
Symbols are as in Figure 4.}
\label{f:fig12}
\end{figure}
\end{center}

\subsection{Barium}

The significant correlation between the abundances of proton and
$n-$capture processing might be an indication of a contribution by thermally
pulsating stars of moderate mass to the nucleosynthesis required to explain the
observed Na-O anticorrelation. Thermally pulsating stars should also produce
primary CNO elements. A star-to-star variation in the sum of CNO elements might be revealed as
a broadening or splitting in the subgiant branch. Such a splitting has been observed
in a few globular clusters, including 47~Tuc but not M~5 (see Piotto et al. 2012; note
however that such a splitting might also be explained as due to an age spread).

Alves-Brito et al. (2005) did not find any sign of
a correlation between Ba and Na abundances from high-dispersion spectra of
four RGB and one RHB star in 47 Tuc. A similar lack of correlation was obtained by D'Orazi
et al. (2010) from moderately high resolution spectra of a large number of
RGB stars, and by Worley and Cottrell (2012) from moderate-resolution 
spectra of RGB stars. However, these samples are either very limited or based
on data of insufficient quality to reveal subtle trends and therefore cannot be 
considered as conclusive.

The only line of an $n-$capture element detectable
in our spectra of RHB stars is the Ba\ts{II} line at 5853.69~\AA, which is usually 
considered a good diagnostic with a negligible hyperfine structure and small
non-LTE corrections (Mashonkina \& Zhao 2006). In the Sun, Ba is predominantly 
produced by the $s-$process (Kappeler et al. 1989). The same likely occurs in 47~Tuc (Brown \& Wallerstein 1992; 
James et al. 2004; Alves-Brito et al. 2005; Wylie et al. 2006), while the low 
Ba/Eu ratio of M~5 stars suggests that for this cluster Ba is mainly produced by
the $r-$process (Ivans et al. 1999, 2001; Yong et al. 2008). 

Figure~\ref{f:fig12} displays the run of [Ba/Fe] with [Na/O] for stars in 47~Tuc.
Once we eliminate the C-star \#81468 (see below), there is a weak but significant 
correlation between these two quantities; the Pearson linear correlation coefficient 
is $r=0.350$\ over 106 stars, which is significant at a very high level of confidence.
A slightly higher and consequently more significant value of $r=0.390$\ is obtained for the 
Spearman rank coefficient. However, we cannot entirely
exclude some analysis effect, because [Na/O] is correlated with temperature along
the HB, and the total range in [Ba/Fe] is only $\sim 0.2$~dex. A confirmation
of this result is necessary before it can be considered definitive proof of
a real correlation between the abundances of $p-$\ and $n-$capture elements in 47~Tuc.


Since there is no observation of a splitting of the subgiant
branch, we did not expect and indeed did not find any correlation
between [Ba/Fe] and [Na/O] for M~5.

\begin{center}
\begin{figure}
\includegraphics[width=8.8cm]{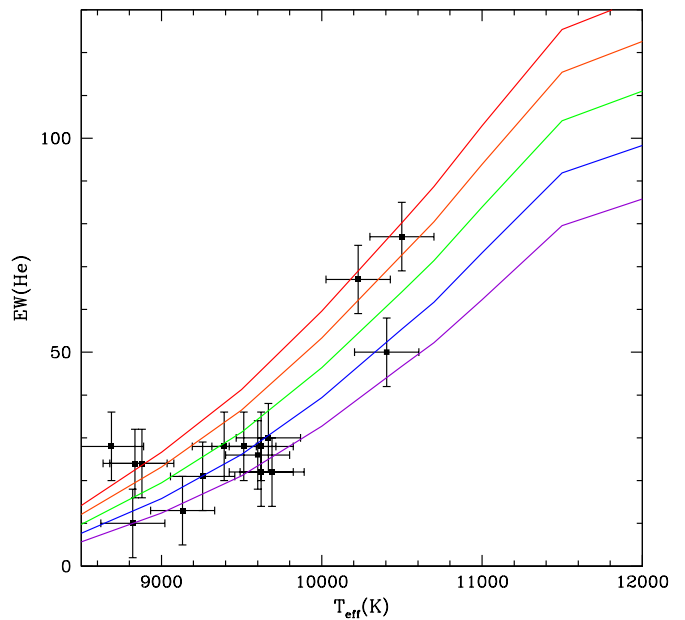}
\caption{Run of the equivalent width of the 5876~\AA\ He\ts{I} feature as a function of
$T_{\rm eff}$\ for stars on the BHB of M~5. Lines are predictions for different helium
abundances [He/H]=-0.4, -0.2, 0, +0.2, and +0.4, from bottom to top. }
\label{f:fig13}
\end{figure}
\end{center}

\subsection{Helium}

We observed the He\ts{I} feature at 5876~\AA\ (actually, a close multiplet) in the spectra 
of the warmer BHB stars in our sample for M~5. Figure~\ref{f:fig13} displays the run of the 
equivalent width of this feature as a function of temperature, along with typical error
bars of $\pm 8$~m\AA in the equivalent width and $\pm 200$~K in the effective temperatures. 
These values are compared with predictions for different values of the He abundances (here 
expressed in terms of [He/H], with reference to an adopted solar value of 0.1 by number of
atoms). There clearly is a large scatter that prevents determination of a
sensible value for individual stars. However, if we average the values we obtain
for stars with $T_{\rm eff}>9000$~K, we obtain an average of [He/H]=$-0.16\pm 0.08$,
which corresponds to an He abundance in mass of $Y=0.22\pm 0.03$. This value is 
lower than the value of $Y\sim 0.26$\ expected if these stars have a cosmological He 
abundance increased by the (small) effect of the first dredge-up (Sweigart 1987). 
However, the difference is only marginally larger than the observational error and
may be attributed to several possible defects in the analysis. It suggests
that these stars are not significantly enriched in He. This agrees with the moderate
[Na/O] ratio obtained for BHB stars. We repeat that this does not mean that there
is no He-rich population in M~5, rather that this population should
populate the VBHB, and would not be sampled by our observations. This
of course sets a limit to the number of very He-rich stars that are 
possibly present in M~5, since the
VBHB stars constitute 33\% of the total of HB stars. This fraction is lower than
predicted by e.g. D'Antona \& Caloi (2008). 

We measured a quite large equivalent
width of the He line for two of the hottest stars in our sample (\#21765, 
$T_{\rm eff}$=10227~K, and \#26450, $T_{\rm eff}$=10500~K; equivalent widths of 67 and 
77~m\AA, respectively). Taken at face values, these equivalent widths indicate a very high 
He abundance ([He/H]$\sim 0.35$, that is Y=0.47). Star \#26450 is also very O-poor 
([O/Fe]=-0.40), but star \#21765 is quite O-rich ([O/Fe]=0.33). Given the large errors 
in our He abundances for individual stars (about 0.3 dex for these stars, mainly due to
uncertainties in the equivalent widths), we do not attach much value to this occurrence.

A comparison with the very similar analysis we performed in Paper II for
stars in NGC~1851 (where we obtained $Y=0.29\pm 0.05$) suggests that the BHB stars 
observed in that cluster are more He-rich than those in M~5. This agrees 
with expectations based on synthesis of the HB populations (see discussion in Paper II).

\begin{center}
\begin{figure}
\includegraphics[width=8.8cm]{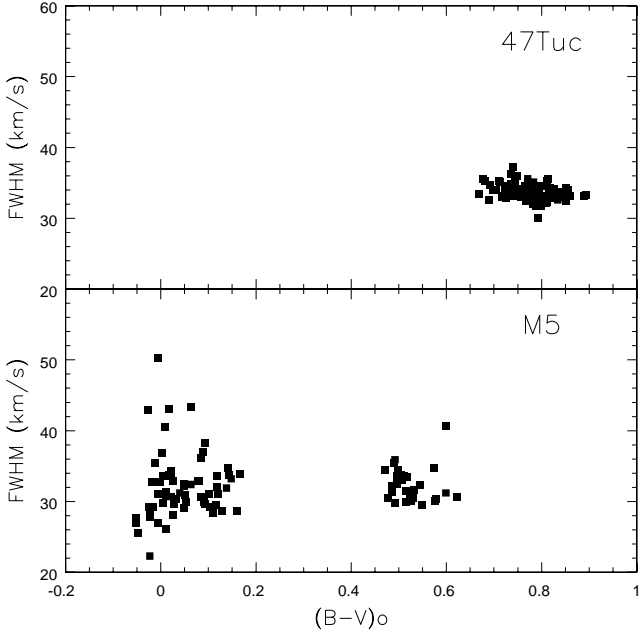}
\caption{Run of the FWHM of the cross-correlation profiles with dereddened $(B-V)_0$\ colours
for 47 Tuc and M~5 (upper and lower panel, respectively).}
\label{f:fwhm}
\end{figure}
\end{center}

\begin{center}
\begin{figure}
\includegraphics[width=8.8cm]{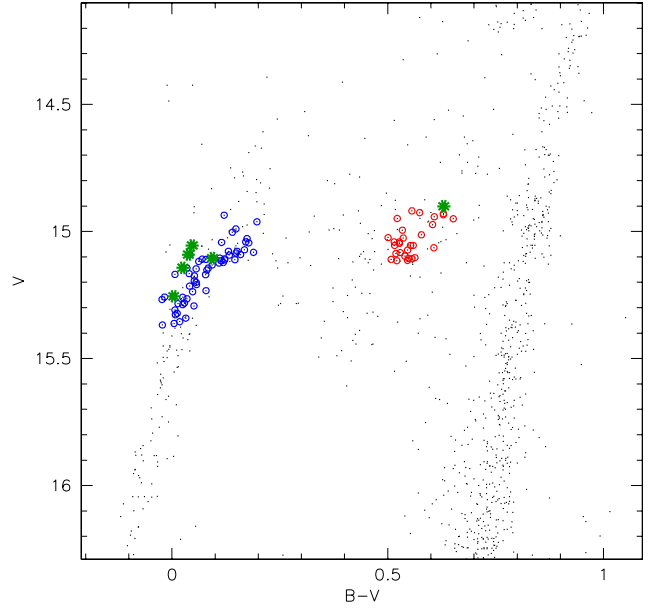}
\caption{Colour-magnitude diagram for M~5. Blue and red open symbols are slow-rotator BHB and RHB
stars, respectively. Green asterisks are fast rotators. Dots are stars that were not observed.}
\label{f:fig14}
\end{figure}
\end{center}

\subsection{Full width at half maximum and rotation}

A number of HB stars are known to be fast rotators (see discussion in Cort\'es et al. 2009).
We examined the full width at half maximum (FWHM) of the cross-correlation profile for evidence of rotation.
Values of the FWHM are listed in Tables~\ref{t:tab1} and \ref{t:tab2}. The run
of the FWHM with dereddened $(B-V)_0$\ colours is plotted in Figure~\ref{f:fwhm}.

In 47~Tuc no star has abnormally high values of the FWHM that would indicate a fast rotation.
However, there is a strong positive correlation between the FWHM and the temperature of the stars. 
Also, stars of the bright group have systematically higher values of the FWHM.
Once deconvolved for the widening caused by the width of the template lines and
the instrumental profile (16.1~km$\ts $s$^{-1}$), the {\it intrinsic} FWHM values range
from 15 km$\ts $s$^{-1}$ for the cooler stars to 20 km$\ts $s$^{-1}$ for the warmer ones, with the
bright group stars having a FWHM about 3 km$\ts $s$^{-1}$ higher than the faint group
ones. We attribute this effect to star-to-star variations of the macroturbulence 
velocity. The values we found are higher than but not incompatible with, those found by Carney
et al. (2008a), who also found a strong correlation between macroturbulence velocity 
and effective temperature among field RHB stars.

In M~5, similar trends (though noisier) are observed for RHB stars, while for BHB 
stars, the FWHM decreases with increasing temperature, and once appropriate 
deconvolution for template and instrumental profile are made, it becomes 
negligible for the warmer stars ($T_{\rm eff}$ warmer than $\sim 9500$~K),
whose atmospheres are expected to be in radiative equilibrium. In addition, one 
RHB star (\#21869) and five BHB stars (\#15529, \#21180, \#28804, \#31189, and 
\#32246) have much broader lines than the remaining ones and 
are likely fast rotators, with \#21180 having an intrinsic FWHM$\sim 42$~km$\ts $s$^{-1}$ 
and all the other fast-rotating stars with intrinsic FWHM$\sim 30-33$~km$\ts $s$^{-1}$.
While fast rotators are common among BHB stars (Cort\'es et al. 2009), they are much 
rarer among RHB ones. Among field RHB stars, the record holder is HD195636, which 
rotates at $V_{\rm rot}~\sin i=22.2\pm 1.0$~km$\ts $s$^{-1}$ (Carney et al. 2008b). While we did 
not calibrate line broadening in terms of rotational velocity, we estimate that \#21869 in M~5 likely has 
a rotational velocity similar to HD195636. Star \#21869 possibly has a variable radial
velocity: there is a 3.2 km$\ts $s$^{-1}$ offset between measures made from gratings HR12 and HR19,
which were taken one week apart, while the average offset for RHB stars is 0.4 km$\ts $s$^{-1}$
with an r.m.s. of 0.56 km$\ts $s$^{-1}$. This star then deviates five times from the typical r.m.s..
In addition, \#21869 is the most Ba-rich star of our sample, and is both Na and O-rich
(but not C-rich). We therefore suspect that \#21869 is a spectroscopic binary and that its 
companion is probably a white dwarf, progeny of the evolution of a rather
massive AGB star. It is interesting to note that \#21869 is the brightest and
one of the reddest among the observed RHB stars of M~5.

We also note that four out of five of the 
BHB fast-rotators are overluminous with respect to the remaining blue HB stars (see Figure~\ref{f:fig14}). 
We found a similar result for the only likely fast rotator found in NGC~1851
(Paper II). No similar correlation between rotation and evolution off the
ZAHB was noticed so far (Cort\'es et al. 2009). This point will be re-examined
in a future paper. 

\subsection{Star \#81468: A C-star on the HB of 47~Tuc }

Star \#81486 (the reddest and among the brightest stars observed in 47~Tuc) is
a C-star. The spectrum is very rich of CN-lines (see Figure~\ref{f:fig3}). The star is out of the mean
loci of the O-Na anticorrelation defined by the other stars, being O-rich
for its Na abundance. It is also very Ba-rich ([Ba/Fe]=1.48). The most likely explanation
of its composition is pollution by a previous low-mass AGB star (currently a white dwarf)
in a binary system.
However, these systems typically have periods around $\sim 1000$~days, hence
we do not expect to find variations of radial velocities because the spectra with 
gratings HR12 and HR19 of 47~Tuc were taken on the same night. Lines are not 
wider than those of other stars in 47~Tuc and the radial velocity is coincident 
with the average of the cluster.


\section{Conclusions}

We presented an analysis of the composition of large samples of stars on the HB 
of 47~Tucanae and M~5. For the first cluster, which is the prototype of metal-rich
GCs with only RHB stars, there are only limited selection effects. For the
second we limited our analysis to the RHB and to the BHB stars (in our
nomenclature the latter includes stars warmer than the RR Lyraes but cooler than the Grundahl et al.
1999 jump). We excluded from our sample RR Lyrae variables and stars warmer than 
11,000~K (in our nomenclature, VBHB stars), for which our analysis would not have
provided reliable results. These two sections of the HB include less than half of
the HB stars in M~5, so that appropriate care must be taken when examining results.
 
In 47~Tuc we found tight relations between colours of the stars and their abundances 
of $p-$capture elements. We can predict quite accurately (within 0.016~mag in $B-V$) 
the colour of an HB star in our faint sample (that is, in the restricted magnitude range
$13.94<V<14.03$) from its [Na/O] ratio. Similar good correlations 
exist for other elements, including N and Al. This strongly supports the idea that the 
He content - which is expected to be closely correlated with the abundances of 
$p-$capture elements - is the third parameter (in addition to overall metallicity and 
age) that determines the colours of HB stars. This agrees with the results of Nataf et al. 
(2011), who used this argument to discuss the radial segregation of first- and 
second-generation stars. Synthetic HB simulations performed for this specific project show that 
the He abundance spread
among the HB stars in 47Tuc is likely small (${\rm \Delta{Y}<0.03}$). This result provides
plain support to the analysis performed by Di Criscienzo et al. (2010) and Milone et al. (2012). If 
we transform the spread in colour of HB stars at a given [Na/O] in terms of mass of 
the stars, the scatter is only 0.016~$M_\odot$\ once it is deconvolved for the (quite small) 
observational errors. This leaves only limited room for additional factors to influence mass 
loss along the RGB, at least for this cluster. 

There possibly is a correlation between the abundances of $p-$\ and $n-$capture 
elements in 47~Tuc, although the total range in Ba abundances is small ($\sim 0.2$~dex)
and there is the chance that this trend is due to analysis errors. If 
confirmed by a more accurate analysis, this might suggest that AGB stars of moderate mass
contributed to the gas from which second generation stars formed in this cluster.

Considering the selection effects in our sample is important to understand the 
results for M~5. In this case, we find that, as expected, RHB stars are Na-poor and O-rich, 
and therefore likely belong to the primordial population. There is a clear correlation of the 
[Na/O] ratio and N abundances with colour along the BHB. A derivation of the He abundance 
for stars in the temperature range $9000<T_{\rm eff}<K$\ yields a low value of 
$Y=0.22\pm 0.03$. This is expected, since HB stars of a putative He-rich population 
in this cluster should be warmer than 11000~K, and are accordingly not sampled by our analysis.
However, while an overall correlation between abundances of $p-$capture elements and 
colours exists for star along the HB of M~5, our data suggest that there should be
an additional source of scatter in the total mass loss of RGB stars, at a level of
$\sim 0.03~M_\odot$; this is more than the scatter observed for the 47~Tuc stars.
Additional observations (e.g. rotational velocities from higher resolution
spectra than used in this paper) are required to understand the mechanism responsible
for this additional scatter.

On the whole, results for these two clusters conform very well to the paradigm that
the distribution of stars along the HB of a GC is determined by their initial He content
and is therefore strongly related to the multiple population phenomenon. They also show
that this comparison requires some care in examining the results for individual
clusters.

We also note that there are six fast rotators on the HB of M~5. Fast rotators are
known to be present among HB stars of several GCs, most of them being on the BHB (Cort\'es
et al. 2009). 
The fraction of fast rotating BHB stars strongly varies from cluster to cluster.
In addition, both in M~5 and in NGC~1851 (considered in Paper II), we found that 
fast-rotating stars are overluminous with respect to other BHB stars; this also might be
different from what was found in other GCs. We will examine these questions in more detail in
a forthcoming paper.

Finally, we found a C-star on the HB of 47~Tuc and a Ba-rich, fast-rotating, likely binary star 
on the HB of M~5. The unusual composition of these stars is probably due to mass transfer from
thermally pulsing AGB stars, which are now white dwarfs. We note that these two stars are 
among the brightest and coolest HB stars.

\begin{acknowledgements}
This publication makes use of data products from
the Two Micron All Sky Survey, which is a joint project of the
University of Massachusetts and the Infrared Processing and Analysis
Center/California Institute of Technology, funded by the National
Aeronautics and Space Administration and the National Science
Foundation. This research has made use of the NASA's Astrophysical
Data System. This research has been funded by PRIN INAF
"Formation and Early Evolution of Massive Star Clusters".
We thank the anonymous referee for a careful revision of the
manuscript and many suggestions that improved the paper.

\end{acknowledgements}

\end{document}